\documentclass[aip,reprint,amsmath,amssymb,]{revtex4-1}

\newif\ifpdf
        \ifx\pdfoutput\undefined
        \pdffalse 
        \else
        \pdfoutput=1 
        \pdftrue
        \fi
\ifpdf
   \usepackage[pdftex]{graphicx}
   \DeclareGraphicsExtensions{.pdf, .png, .jpg}
   \graphicspath{{./FIG/}}
   \usepackage{thumbpdf}
\else
   \usepackage{graphicx}
   \DeclareGraphicsExtensions{.eps}
   \graphicspath{{./FIG/}}
\fi

\usepackage{bm}
\usepackage{dcolumn}
\usepackage[yyyymmdd,hhmmss]{datetime}
\usepackage[colorlinks=true,urlcolor=blue,citecolor=blue,linkcolor=blue,breaklinks=true]{hyperref}
\usepackage{orcidlink}

\begin{document}

\title{Computing the Frequency-Dependent NMR Relaxation of $^1$H Nuclei in Liquid Water}

\author{Dietmar Paschek\orcidlink{0000-0002-0342-324X}}
\email{dietmar.paschek@uni-rostock.de}
\affiliation{Institut f\"ur Chemie, Abteilung Physikalische und Theoretische Chemie, 
Universit\"at Rostock, Albert-Einstein-Str.~27, D-18059 Rostock, Germany}

\author{Johanna Busch\orcidlink{0000-0003-3784-0188}}
\affiliation{Institut f\"ur Chemie, Abteilung Physikalische und Theoretische Chemie, 
Universit\"at Rostock, Albert-Einstein-Str.~27, D-18059 Rostock, Germany}

\author{Eduard Mock}
\affiliation{Institut f\"ur Chemie, Abteilung Physikalische und Theoretische Chemie, 
Universit\"at Rostock, Albert-Einstein-Str.~27, D-18059 Rostock, Germany}

\author{Ralf Ludwig\orcidlink{0000-0002-8549-071X}}
\affiliation{Institut f\"ur Chemie, Abteilung Physikalische und Theoretische Chemie, 
Universit\"at Rostock, Albert-Einstein-Str.~27, D-18059 Rostock, Germany}
\affiliation{Leibniz-Institut f\"ur Katalyse an der Universit\"at Rostock e.V., Albert-Einstein-Str.~29a, D-18059 Rostock, Germany}
\affiliation{Department Life, Light \& Matter, Universit\"at Rostock, 
Albert-Einstein-Str.~25, D-18059 Rostock, Germany}

\author{Anne Strate\orcidlink{0000-0001-6621-8465}}
\email{anne.strate@uni-rostock.de}
\affiliation{Institut f\"ur Chemie, Abteilung Physikalische und Theoretische Chemie, 
Universit\"at Rostock, Albert-Einstein-Str.~27, D-18059 Rostock, Germany}

\date{\today~at~\currenttime}

\begin{abstract}
It is the purpose of this paper to present a 
computational framework for
reliably
determining the frequency-dependent 
intermolecular and intramolecular
NMR dipole-dipole relaxation rate
of spin $1/2$ nuclei
from MD simulations. 
The approach avoids alterations caused by  
well-known
finite-size effects
of the translational diffusion. Moreover, a procedure
is derived to control and correct for 
effects caused by fixed distance-sampling cutoffs and
periodic boundary conditions.
By construction,  this approach is capable of accurately predicting the 
correct low-frequency scaling behavior
of the intermolecular  NMR dipole-dipole relaxation rate and
thus allows the reliable calculation of the
frequency-dependent relaxation rate over many orders of magnitude.
Our approach is based on the utilisation of the theory
of Hwang and Freed 
for the 
intermolecular
dipole-dipole correlation function and its corresponding
spectral density [J. Chem. Phys. \textbf{63}, 4017 (1975)]
and its combination with data from molecular dynamics (MD) simulations.
The deviations from the Hwang and Freed theory
caused by periodic boundary conditions and
sampling distance cutoffs 
are quantified by means of random walker
Monte Carlo simulations. 
An expression based on the Hwang and Freed theory
is also suggested for
correcting those effects. 
As a proof of principle, our approach is demonstrated by computing the
frequency-dependent inter- and intramolecular dipolar NMR
relaxation rate of the
$^1$H nuclei in liquid water 
at $273\,\mbox{K}$ and $298\,\mbox{K}$
based on
simulations of
the TIP4P/2005 model. Our calculations are suggesting that 
the intermolecular contribution to the $^1$H NMR relaxation 
rate of the TIP4P/2005 model 
in the extreme narrowing limit
has previously been substantially
underestimated.
\end{abstract}

\keywords{NMR, NMR Relaxation, Molecular Dynamics Simulations, Water, Diffusion}

\maketitle

\section{Introduction}

The primary mechanism for 
relaxation of spin $1/2$ nuclei in NMR spectroscopy is based on 
their magnetic 
dipole-dipole interactions which are mediated by 
intermolecular and intramolecular motions.\cite{Abragam1961,kowalewski_2013} 
Frequency-dependent NMR
relaxation data can be used
to provide an understanding of the details of molecular 
motion within a chemical system.\cite{Kruk_2011,Kruk_2012} 
However, interpreting experimental data often
requires models that are specific to certain systems 
and/or conditions and assume analytical forms of the
relevant time correlation functions.\cite{Sholl_1981,Hwang_1975} 
Since these models may not account for all molecular-level dynamical
processes, it can be sometimes difficult to assess whether a 
certain model is appropriately
describing a particular system.\cite{overbeck_2020,overbeck_2021}
To address these limitations, 
Molecular Dynamics (MD) simulations can be used to study NMR relaxation phenomena 
using a ``first principles''-based approach
without the need for analytical models. 
Hence, the value of MD simulations in interpreting NMR relaxation data has 
been recognized from early on.\cite{Westlund:1987,schnitker_1987}

Since dipolar NMR relaxation is due to the fluctuating fields resulting from the 
magnetic dipole-dipole
interaction 
between two spins, formally a division into both, intermolecular
and intramolecular 
contributions can be performed. Here intramolecular dipolar relaxation is driven by molecular 
vibrations, conformational changes, and rotations. Intermolecular contributions,
on the other hand, are 
primarily driven by translational diffusion. 
They are, however, also affected by
librations, conformational changes and rotational motions
on short time scales.
Considering the complexity of this convolution of dynamical phenomena,
it can be quite challenging to disentangle all their different contributions.

Moreover, 
the accurate computation of intermolecular contributions 
to the relaxation rate from MD simulations poses serious challenges.
Since the relaxation rate largely depends on translational diffusion, the
exact size of the 
self-diffusion coefficients matters.
Diffusion coefficients
obtained from MD simulations with periodic boundary conditions, however,
 are known to exhibit a non-negligible system size dependence \cite{yeh_2004,moultos_2016,busch_2023c}. 
 Hence the computed intermolecular relaxation rates
are also system-size dependent.
Another important influence of the system size on the computed spectral densities
has been recently pointed out by Honegger et al. \cite{honegger_2021}, suggesting
that an accurate representation of the low frequency requires properly
covering long intermolecular distance ranges.
In addition to that, the accurate representation of the low-frequency limiting behavior
of the relaxation rate will also require very long simulations, covering nearly
``macroscopic'' time scales. 
Hence, the accurate computation of intermolecular
relaxation rates is twofold burdened by having to consider simulations
of large systems for very long times.

To deal with both problems, we present a 
computational framework designed to
determine the frequency-dependent 
intermolecular
NMR dipole-dipole relaxation rate
from MD simulations. Our approach is based on a separation
of the intermolecular part into
a purely diffusion-based component, which is represented
by the theory of Hwang and Freed \cite{Hwang_1975}, and another component,
which contains the difference between
the Hwang and Freed model and the correlation functions computed
from MD simulations. It is shown, that for long times the
second term 
effectively decays to zero and thus exhibits an inherent short-term nature.
Hence, by construction,  this approach is capable of accurately predicting the 
correct low-frequency scaling behavior
of the intermolecular  NMR dipole-dipole relaxation rate.
System-size dependent diffusion coefficients can be dealt with
by employing Yeh-Hummer \cite{yeh_2004,moultos_2016} corrected inter-diffusion 
coefficients for the Hwang and Freed model.
Additional deviations 
caused by periodic boundary conditions and
limited sampling distance cutoffs 
are thoroughly studied by means of random walker
Monte Carlo simulations of the Hwang and Freed model. 
Moreover, we show that
the theory by Hwang and Freed can also be utilised, to some extent, to
correct for those effects as well. 

Our approach is demonstrated by computing the
frequency-dependent 
intermolecular and intramolecular dipolar NMR
relaxation rates of the
$^1$H nuclei in liquid water 
at $273\,\mbox{K}$ and $298\,\mbox{K}$
based on
simulations of
the TIP4P/2005 model for water \cite{abascal_2005}. Our calculations suggest that 
the intermolecular contribution to the $^1$H relaxation 
rate of the TIP4P/2005 model 
in the extreme narrowing limit
has previously been 
underestimated \cite{Calero_2015}.

\section{Theory: Dipolar NMR Relaxation and Correlations in the 
Structure and Dynamics of Molecular Liquids}

The dipolar relaxation rate of an NMR active nucleus is 
determined by its magnetic dipolar interaction
with all the surrounding nuclei.
It is therefore subject to the time-dependent spatial correlations
in the liquid
and is  affected by
both the molecular structure and the dynamics of the liquid.
For the NMR relaxation rate of nuclear
spins with $I = 1/2$, the magnetic dipole-dipole interaction
represents the dominant contribution.\cite{Abragam1961} 
The frequency-dependent relaxation rate, i.e. the rate at which the nuclear spin
system approaches thermal equilibrium, is determined by the time dependence
of the magnetic dipole-dipole coupling.
For two like spins, it is given by \cite{Abragam1961,Westlund:1987}
\begin{eqnarray}\label{eq:relax}
R_1(\omega) & = &
\gamma^4\hbar^2I(I+1)(\mu_0/4\pi)^2\times
\\
& &
\left\{
\int\limits_{-\infty}^\infty 
\left< \sum_j^N
\frac{D_{0,1}[\Omega_{ij}(0)]}{r_{ij}^3(0)}
\cdot
\frac{D_{0,1}[\Omega_{ij}(t)]}{r_{ij}^3(t)}
\right> e^{i\omega t} dt
\right. \nonumber\\
&+&4\left.
\int\limits_{-\infty}^\infty 
\left< \sum_j^N
\frac{D_{0,2}[\Omega_{ij}(0)]}{r_{ij}^3(0)}
\cdot
\frac{D_{0,2}[\Omega_{ij}(t)]}{r_{ij}^3(t)}
\right> e^{i2\omega t} dt
\right\}, \nonumber
\end{eqnarray}
where $D_{k,m}[\Omega]$ is the $k,m$-Wigner 
rotation matrix element of rank $2$.
The Eulerian angles $\Omega(0)$ and $\Omega(t)$ at time zero and time $t$
specify the dipole-dipole vector relative to the laboratory fixed frame
of a pair of spins and $r_{ij}$ denotes their separation distance and
$\mu_0$ specifies the permeability of free space.
The sum indicates the summation of all $j$ interacting like spins
in the entire system.
In case of an isotropic
fluid both spectral densities in Equation \ref{eq:relax}
are represented by the same function \cite{Westlund:1987}
\begin{eqnarray}
\label{eq:jomega_1}
J(\omega) = \frac{2}{5}\;\mathrm{Re}\left\{ \int\limits_0^\infty 
G(t) \;e^{i\omega t} dt\right\}
\end{eqnarray}
where $G(t)$ denotes
the ``dipole-dipole correlation function''
which is available via
\cite{Westlund:1987,Odelius:1993}
\begin{eqnarray}\label{eq:dipolcor}
G(t) & = & \left< \sum_{j} r_{ij}^{-3}(0)\,r_{ij}^{-3}(t)
P_2\left[\,\cos \theta_{ij}(t) \right] \right>\;,
\end{eqnarray}
where
$\cos \theta_{ij}(t)$ is the cosine of the
angle between the connecting vectors $\vec{r}_{ij}$
joining spins $i$
and $j$ at time $0$ and at time $t$ while
$P_2[\ldots]$ represents the second Legendre polynomial.\cite{Westlund:1987} 
Given the case of rotational isotropy,
Equation \ref{eq:dipolcor} results from Equation \ref{eq:relax} by
aligning the magnetic field vector $\vec{B}$ with the orientation of
the connecting vector $\vec{r}_{ij}(0)$, thus allowing for a more
efficient sampling of the angular contributions. Integrating
over all field vector orientations then results in a pre-factor of 1/5.

By combining Equations \ref{eq:jomega_1} and \ref{eq:dipolcor},
the spectral density 
\begin{eqnarray}\label{eq:jomega}
J(\omega)&=&
\frac{2}{5}
\left<
  \sum_j r^{-6}_{ij}(0)
\right> 
\mathrm{Re}\left\{
\int\limits_0^\infty G^\mathrm{n}(t) \;e^{i\omega t} dt
\right\}
\end{eqnarray}
can be expressed as being composed of
a $r_{ij}^{-6}$ averaged 
constant
containing solely structural information and the Fourier-transform
of a normalized correlation function $G^\mathrm{n}(t)=G(t)/G(0)$, 
which is sensitive to the motions of
the molecules within the liquid.

For the case of the extreme narrowing limit $\omega\rightarrow 0$
we obtain a relaxation rate 
\begin{eqnarray}
R_{1}(0)
& = &
\gamma^4 \hbar^2
I(I+1)
\left(\frac{\mu_0}{4\pi}\right)^2
\cdot 2\int\limits_0^\infty
G(t) \,dt\;,
\end{eqnarray}
where the integral over the dipole-dipole correlation function 
\begin{eqnarray}\label{eq:ddcorel}
\int\limits_0^\infty G(t) \,dt & = & \left<
  \sum_j r^{-6}_{ij}(0)
\right> \tau_G\;
\end{eqnarray}
is the product of the $r_{ij}^{-6}$ averaged constant
and a correlation time $\tau_G$, which is the time-integral
of the normalized correlation function 
\begin{equation}
\tau_G=
\int\limits_0^\infty
G^\mathrm{n}(t) \,dt\;.
\end{equation}
Here $\tau_G$ represents the dynamical
contributions from the time correlations
of the molecular motions
within the liquid.

The correlation function $G(t)$, and hence 
$J(\omega)$ and
$R_1(\omega)$
can be calculated directly from
MD-simulation trajectory data. However, as we will show later,
the computed correlation functions are subject to 
system size effects and
the way how periodic boundary conditions are treated.

From the definition of the dipole-dipole correlation function in 
Equation \ref{eq:dipolcor} follows directly
that the relaxation rate $R_1(\omega)$
is affected  by internal, reorientational and translational
motions in the liquid. Moreover, it is obvious that it also
depends strongly on the average 
distance between the spins
and is hence sensitive to changing intermolecular and intramolecular
pair distribution functions \cite{Hertz1967, Hertz1976}. 
In addition, the $r_{ij}^{-6}$-weighting introduces a particular
sensitivity to changes occurring at short distances.
For convenience, one may divide the spins $j$ into different
classes according to whether they belong to the same molecule as
spin $i$, or not, thus arriving at an {\em inter}- and {\em
intramolecular} contribution to the relaxation rate
\begin{eqnarray}
R_1(\omega) = R_{1,\rm inter}(\omega) + R_{1,\rm intra}(\omega) ,
\end{eqnarray}
which are determined by corresponding 
intermolecular and 
intramolecular
dipole-dipole correlation functions
$G_\mathrm{intra}(t)$ and $G_\mathrm{inter}(t)$.
The  intramolecular contribution is basically due
to molecular reorientations and conformational changes
and has been used extensively to study the reorientational
motions, such as that of the H-H-vector in $\mbox{CH}_3$-groups in
molecular liquids and crystals \cite{Stejksal:1959}.
The intermolecular contributions are mostly affected by the translational
mobility (i.e. diffusion) within the liquid and the preferential
aggregation or interaction between particular sites, as expressed
by intermolecular pair correlation functions.

\paragraph{Intermolecular Contributions}

The structure of the liquid can be expressed in terms
of the intermolecular site-site pair correlation function $g_{ij}(r)$, 
describing the probability of finding
a second atom of type $j$ in a distance $r$ from a reference site 
of type $i$ according to \cite{Egelstaff}
\begin{equation}
g_{ij}(r) = \frac{1}{N_i\,\rho_j} 
\left< 
\sum_{k=1}^{N_i}
\sum_{l=1}^{N_j}
\delta(\vec{r}-\vec{r}_{kl})
\right>\;,
\end{equation}
where $\rho_j$ is the number density of spins of type $j$.
The pre-factor  of the intermolecular dipole-dipole correlation
function is hence related to the pair distribution function via
an $r^{-6}$-weighted integral over the pair correlation function
\begin{equation}\label{eq:r6gr}
\left<\sum_j r^{-6}_{ij}(0)\right> = \rho_j \,4\pi
\int\limits_0^\infty r^{-6} \; g_{ij}(r)\; r^2\,dr\;.
\end{equation}
Since the process of association in a molecular system
is equivalent to an increasing nearest neighbor peak
in the radial distribution function, Equation \ref{eq:r6gr} 
establishes a quantitative relationship between the degree of
intermolecular association and the intermolecular dipolar
nuclear magnetic relaxation rate.

The integral in Equation \ref{eq:r6gr}, of course, contains all
the structural correlations affecting the spin pairs.
Averaged intermolecular distances between two spins
$\alpha$ and $\beta$ are represented by the integral
\begin{equation}
  \label{eq:dhh-1}
  I_{\alpha\beta}=4\pi\int\limits_0^\infty r^{-6}\,g_{\alpha\beta}(r)\,r^2 dr\;.
\end{equation}
Relating the structure of the liquid to a structureless hard-sphere fluid, 
the size of the integral
$I_{\alpha\beta}$ is conviently described by a ``distance of closest approach''
$d_{\alpha\beta}$, which represents an integral of the
same size, but over a step-like 
unstructured pair correlation function according to
\begin{equation}
  I_{\alpha\beta}=4\pi\int\limits_{ d_{\alpha\beta}}^\infty r^{-6}\cdot 1\cdot r^2 dr=
  \frac{4\pi}{3}\cdot\frac{1}{d_{\alpha\beta}^3}\;.
\end{equation}
Hence the ``distance of closest approach'' can be determined with
the knowledge of $I_{\alpha\beta}$ as 
\begin{equation}
  \label{eq:dhh-2}
  d_{\alpha\beta}=\left[\frac{4\pi}{3}\cdot\frac{1}{I_{\alpha\beta}}\right]^{1/3}\;.
\end{equation}
It is typically assumed that this ``distance of closest approach'' is
identical to the distance used in the structureless hard-sphere diffusion model
as outlined by Freed and Hwang.\cite{Hwang_1975} To determine the
distance of closest approach in
this paper, the integral $I_{\alpha\beta}$ is evaluated 
by integrating over the pair correlation function
numerically up to
half of the box-length $L/2$ and then  corrected by adding the term
$32\pi/(3L^3)$ as long-range correction.

\paragraph{Intramolecular Contributions:}

Intramolecular correlations are computed directly over all involved spin pairs
of type $\alpha$ and $\beta$.
 
Here $\delta_{ij}$ ensures that contributions from identical spins
for the case of $\alpha\!=\!\beta$ are not
counted. Note that for the special case of $\alpha\!=\!\beta$ 
the normalisation has to be modified accordingly: $N_\beta=N_\alpha-1$.
In the case of the water molecule, there is only one intramolecular 
dipole-dipole interaction with a fixed H-H distance when using
a rigid water model such as TIP4P/2005. The intramolecular contribution
to the relaxation rate is therefore solely based on 
the reorientation of the intramolecular H-H vector.
\begin{figure}
        \includegraphics[width=0.3\textwidth]{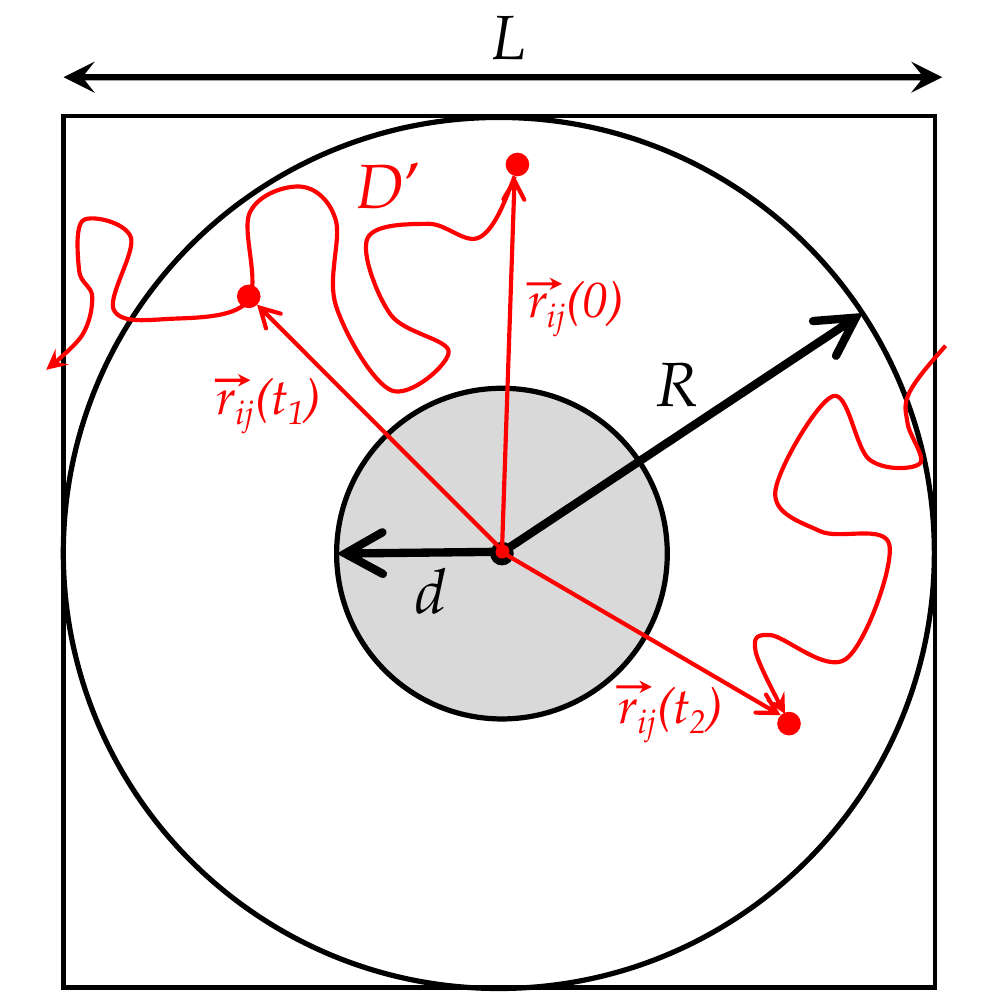}        
         \caption{\label{fig:rwalker} 
         Schematic representation of a
         random walker $j$ travelling with a diffusion coefficient 
         $D'$ within the frame of reference of a particle $i$  
         (located at the origin)
         with and without
         periodic boundary conditions. The shaded volume with radius $d$ 
         (distance of closest approach)
         marks the region avoided by random walker $j$. The sphere with radius
         $R$ indicates the volume, where starting positions 
         $\vec{r}_{j}(0)$ with $r_{ij}(0)\!>\!d$ 
         are sampled from if periodic boundary conditions are not applied. 
         For random walker simulations with periodic boundary conditions, the starting 
         positions are uniformly sampled from the entire box volume with box-size $L$
         applying the condition $r_{ij}(0)\!>\!d$. }
\end{figure}

\section{Methods}

\subsection{Random Walker Monte Carlo Simulations}
\label{sec:randomwalker}

Here we outline our use of random walker 
Monte Carlo simulations exploring 
the diffusion-based contribution to the 
NMR dipolar relaxation
with and
without periodic boundary conditions (PBCs). The random walker simulations without PBCs
are designed to match the conditions of the theory outlined
by Hwang and Freed \cite{Hwang_1975}.
The simulations are carried out within the frame of reference of particle $i$,
which hence stays fixed at the origin of the coordinate system. The diffusion 
coefficient $D'$ therefore represents inter-diffusion coefficients of
both particles $i$ and $j$ with $D'=D_i+D_j$.
As illustrated in \figurename\ \ref{fig:rwalker}, 
if no PBCs are considered,
the starting position of
the random walker $\vec{r}_j(0)$ are sampled from the distance interval
between $d\!<\!r_{ij}(0)\!<\!R$ along the $z$-axis
using a $r^{-2}$ weighting.
To represent a proper volume sampling, 
contributions from 
each individual trajectory are correspondingly weighted 
by a factor $r^{4}_{ij}(0)$. This realization of ``importance sampling''
strongly reduces the statistical noise compared to the unbiased volume sampling,
in particular for large values of $R$.
For simulations with PBCs, however, the starting positions
are sampled uniformly from the volume of the cubic box with box-size $L$
while obeying  the condition $r_{ij}(0)\!>\!d$.
At $t\!=\!0$
each walker starts from its randomly selected starting position $\vec{r}_{j}(0)$.
New coordinates are computed for discrete time intervals 
$\delta t\!=\!10^{-3}$ time units
from
$\vec{r}_j(t+\delta t)=\vec{r}_j(t)+\vec{o}$, where $\vec{o}$ is a vector with random
orientation and $|\vec{o}|=(6D'\delta t)^{1/2}$.
Trial positions that would end up within the spherical volume
 with radius $r_{ij}<d$ are reflected from the sphere and
corrected such that they are compatible with the reflective boundary conditions
used in the theory of Hwang and Freed. If used,
periodic boundary conditions are applied in the sense that
the diffusing particle, when leaving the box on one side, will enter on the 
opposite side, as illustrated in \figurename\ \ref{fig:rwalker}. 
Dipole-dipole correlation functions reported here
are computed by sampling over $10^8$ individual trajectories.

\subsection{MD Simulations}

We have performed MD 
simulations of liquid water using the TIP4P/2005 model \cite{abascal_2005},
which has been demonstrated to rather accurately describe the
properties of water compared to other
simple rigid nonpolarizable water models.\cite{vega_2011}
The simulations are carried out at $273\,\mbox{K}$ and
$298\,\mbox{K}$
under $NVT$ conditions using system-sizes of 512, 1024, 2048, 4096,
and 8192 molecules. The chosen densities correspond to a pressure
of $1\,\mbox{bar}$ at the respective temperatures.
 MD simulations of 1\,ns length each were performed
 using \textsc{Gromacs} 5.0.6.\cite{gromacs4,gromacs3}
The integration time step for all simulations was $2\,\mbox{fs}$.
The temperature of the simulated systems was controlled employing the
Nos\'e-Hoover thermostat~\cite{Nose:1984,Hoover:1985}
with
a coupling time $\tau_T\!=\!1.0\,\mbox{ps}$.
Both, the Lennard-Jones and electrostatic interactions were treated by smooth 
particle mesh Ewald summation.\cite{Essmann:1995,wennberg_2013,wennberg_2015}
The Ewald convergence parameter 
was set to a relative accuracy of the Ewald sum of $10^{-5}$ 
for the Coulomb- and $10^{-3}$ for the LJ-interaction. 
All bond lengths were kept fixed during the simulation run and
distance constraints were solved by means of 
the SETTLE procedure.\cite{miyamoto_1992}

To compute the intermolecular magnetic dipole-dipole correlation functions,
many
autocorrelation functions over relatively large time sets have to be computed
with a high time resolution of $10\,\mbox{fs}$.
To evaluate time 
correlation functions for large time sets with 
$10^5$
entries efficiently, we applied the 
convolution theorem
using fast Fourier transformation (FFT).\cite{grigera_arxiv,numrecipes}
The computations of the properties from MD simulations were done using our
home-built 
software package {\sc MDorado}
based on
the {\sc MDAnalysis} \cite{mdanalysis1,mdanalysis2}, {\sc NumPy} \cite{numpy},
and {\sc SciPy} \cite{scipy} frameworks. {\sc MDorado} is available via Github.

\section{Results and Discussion}

\subsection{Intermolecular Dipole-Dipole Relaxation: Random Walker Monte Carlo Simulations and the Theory of Hwang and Freed}

\begin{figure*}
        \includegraphics[width=0.3\textwidth]{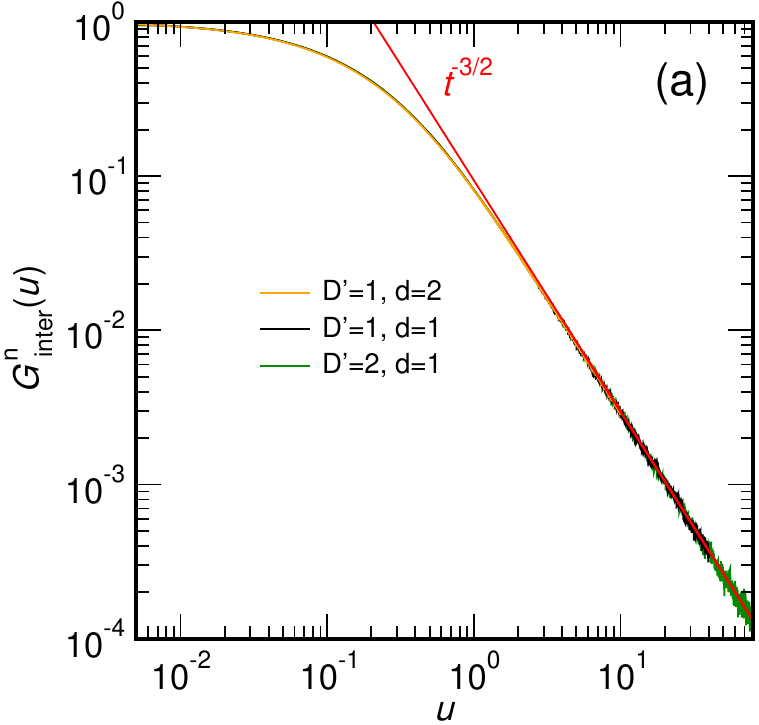}  
        \includegraphics[width=0.3\textwidth]{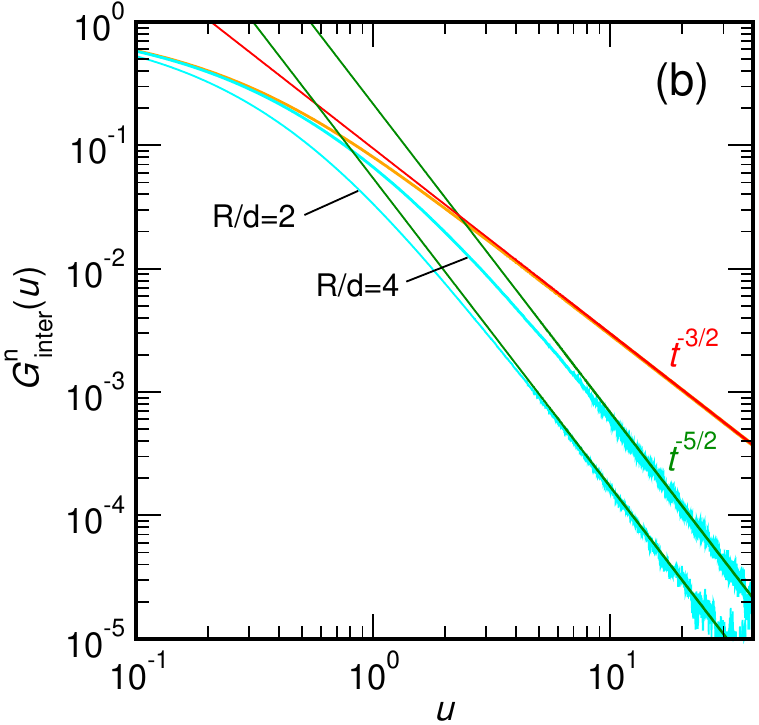}   
        \includegraphics[width=0.3\textwidth]{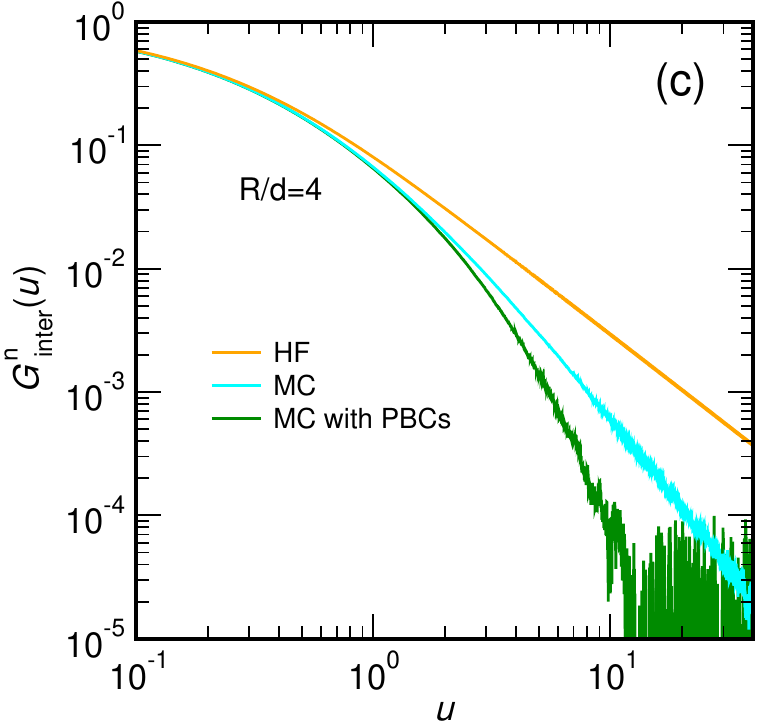}                                   
        \caption{\label{fig:scalemaster} 
        Normalized intermolecular dipole-dipole correlation functions
        $G^\mathrm{n}_\mathrm{inter}(u)$ computed from random walker models
        plotted on a reduced time-scale $u\!=\!D't/d^2$. 
        a)
        $G^\mathrm{n}_\mathrm{inter}(u)$ computed with 
        a sampling cutoff radius $R\!=\!100$ length units
        for different distances of closest approach $d$ and
        inter-diffusion coefficients $D'$.
        Shown in red is also the long-time limiting 
        $\lim_{u\rightarrow\infty}G^\mathrm{n}_\mathrm{inter}(u)\!\propto \!t^{-3/2}$ 
        behavior following Equation
        \ref{eq:Gu32}.
        b)
         The orange line indicates $G^\mathrm{n,HF}_\mathrm{inter}(u)$ 
         according to Equation \ref{eq:g2model_u}.          
         It is showing a $t^{-3/2}$-scaling behavior for long times
         according to Equation \ref{eq:Gu32} indicated in red.
         Applying a cutoff for the starting positions of the random walkers
         affects the computed $G^\mathrm{n,MC}_\mathrm{inter}(u)$ shown in 
         turquoise.
         These functions are depending on the 
         chosen ratio of
         cutoff radius and distance of closest approach $R/d$.
         For long times the 
         $G^\mathrm{n,MC}_\mathrm{inter}(u)$ exhibit a $t^{-5/2}$-scaling 
         shown as green straight lines computed from Equation
         \ref{eq:g2cutofflong}.
         c)
         The orange line indicates $G^\mathrm{n,HF}_\mathrm{inter}(u)$ 
         according to Equation \ref{eq:g2model_u}.
         Applying a cutoff for the starting positions of the random walkers
         affects the computed $G^\mathrm{n,MC}_\mathrm{inter}(u)$ shown in turquoise.
         The curves shown in green represent $G^\mathrm{n,MC}_\mathrm{inter}(u)$-functions
         that were computed applying periodic boundary conditions with starting
         positions sampled from a cubic box with box-length $L$ for comparable
         ratios $R/d=L/(2d)=4$.
        }
\end{figure*}
\begin{figure*}
        \includegraphics[width=0.3\textwidth]{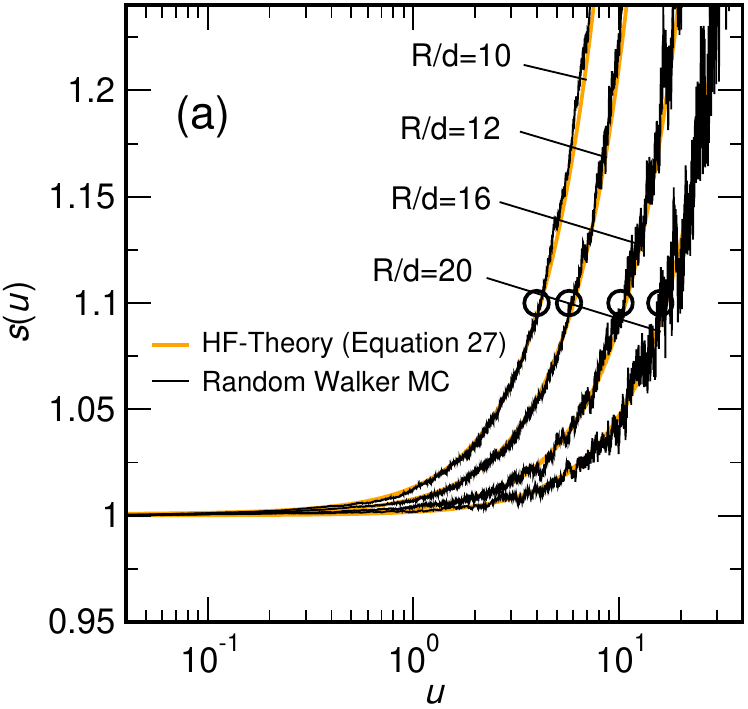}        
        \includegraphics[width=0.3\textwidth]{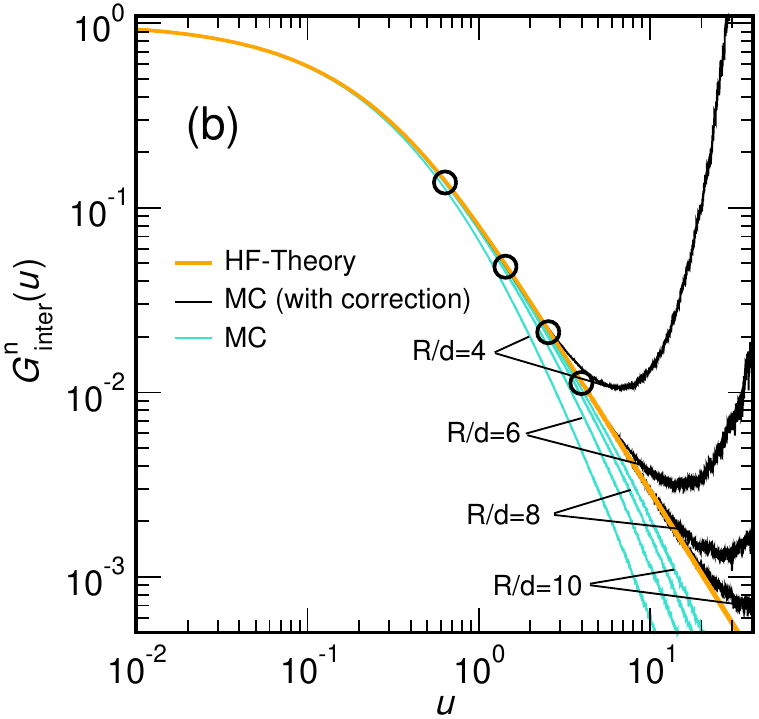}        
        \includegraphics[width=0.3\textwidth]{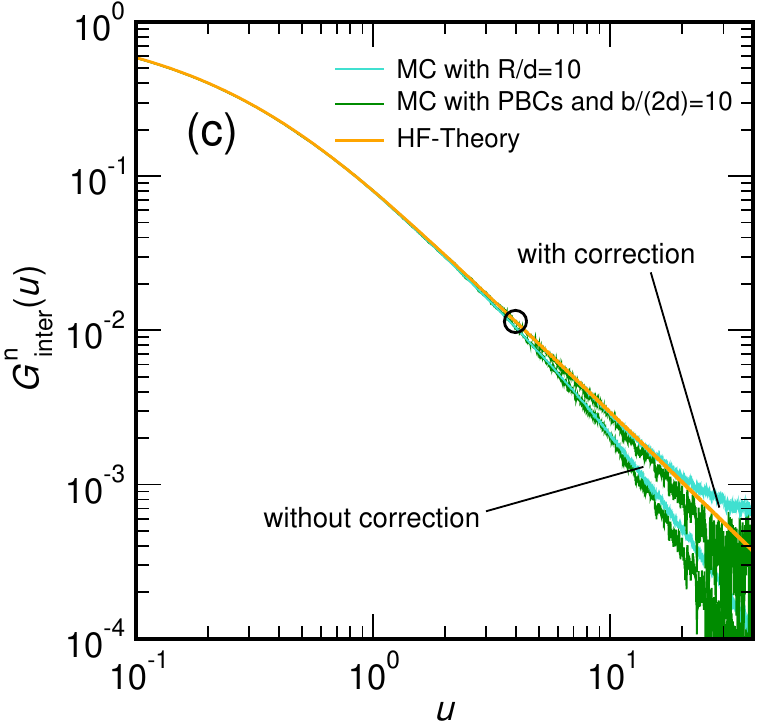}        
        \caption{\label{fig:scale_u} 
        Corrections for
        $G^\mathrm{n}_\mathrm{inter}(u)$ obtained from simulations with
        a fixed sampling cutoff $R$ or periodic boundary conditions.
        Open symbols indicate $G^\mathrm{n,HF}_\mathrm{inter}(u_\mathrm{tr})$ with 
        $u_\mathrm{tr}=R^2/(8\pi d^2)$ or $u_\mathrm{tr}=L^2/(32\pi d^2)$.
        a) Computed scaling functions $s(u)$ for
        various selected ratios $R/d$ 
        employing a lower integration limit of the HF-function
        with $x_\textrm{P}\!=\!\theta\cdot d/R$
        according to Equation \ref{eq:scaling} using
        $\theta\!=\!2.53$.
        b)
        Orange line: 
        $G^\mathrm{n,HF}_\mathrm{inter}(u)$ representing the Hwang-Freed theory.
        Turquoise lines:
        $G^\mathrm{n,MC}_\mathrm{inter}(u)$
        computed from random walker MC simulations for various ratios $R/d$.
        black lines: 
        corrected 
        $G^\mathrm{n}_\mathrm{inter}(u)=G^\mathrm{n,MC}_\mathrm{inter}(u)\cdot 
        s(u,x_P)$.
        c) Normalized  intermolecular
        dipole-dipole correlation functions
        $G^\mathrm{n}_\mathrm{inter}(u)$. Orange line: Hwang-Freed theory
        $G^\mathrm{n,HF}_\mathrm{inter}(u)$. Turquoise lines:
        Data obtained from random walker MC simulations with $R/d=10$ with and without
        correction according to Equation \ref{eq:scaling}.
        Green lines: Data obtained from random walker MC simulations with periodic
        boundary conditions with $L/(2d)=10$ with and without
        correction according to Equation \ref{eq:scaling}.
        }
\end{figure*}

Let us consider the normalized  intermolecular dipole-dipole correlation function
$G^\mathrm{n}_\mathrm{inter}(t)=G_\mathrm{inter}(t)/G_\mathrm{inter}(0)$
computed from the random walker Monte Carlo simulations
 shown in \figurename\  \ref{fig:scalemaster}a
for varying inter-diffusion coefficients $D'$ and distances of closest
approach $d$.
Here we introduce a reduced timescale $u$ based
on $d$ and $D'$ using
\begin{equation}
u\equiv \frac{D't}{d^2}\;.
\end{equation}
Employing the timescale $u$, all $G^\mathrm{n}_\mathrm{inter}(u)$ computed
from random walkers for varying parameters
$D'$ and $d$ collapse on the same curve
as shown in \figurename\ \ref{fig:scalemaster}a.
Following the approach of Hwang and Freed \cite{Hwang_1975}, we can 
also give an 
analytical integral expression for the
dipole-dipole
correlation function  of two
diffusing particles with a distance of closest approach $d$,
reflecting boundary conditions at $r=d$, and
an inter-diffusion coefficient $D'$.
The corresponding normalized correlation function
on the reduced timescale $u$ is given by
\begin{equation}
\label{eq:g2model_u}
G^\mathrm{n,HF}_\mathrm{inter}(u)= \frac{1}{A}\;
\int\limits_0^\infty \frac{x^2\cdot e^{-u\cdot x^2}\;dx}{81+9x^2-2x^4+x^6}\;,
\end{equation}
where $x$ represents a reduced inverse distance scale 
$x\equiv d/r$. 
The constant $A$ follows from computing the integral
for $u\!=\!0$ with
\begin{equation}
A=\int\limits_0^\infty \frac{x^2\;dx}{81+9x^2-2x^4+x^6}
=\frac{\pi}{54}\;.
\end{equation}
For other values of $u$ this integral 
needs 
to be evaluated either
numerically or by using the 
analytically integrated form
derived 
 by Hwang and Freed via partial fractions.\cite{Hwang_1975} 
 For the purpose of our paper, however, 
 the integral form given in Equation
 \ref{eq:g2model_u} turns out to be particularly useful.
 Note also that for short times
$u$ the function $G^\mathrm{n}_\mathrm{inter}(u)$ can be expressed 
via a first-order expansion of the exponential function and is hence
showing a linear time dependence
with
\begin{equation}
\lim_{u\rightarrow 0}G^\mathrm{n,HF}_\mathrm{inter}(u)\approx
1 - 9u\;.
\end{equation}
However, this linear time dependence 
may not properly represent molecular processes in liquids, since the
corresponding time regime will more likely
be dominated by oscillatory 
and jump-like motions. This is in accordance with the observation of
Sholl, who has pointed out that the exact functional form
of the high frequency limit of the 
intermolecular spectral density is highly sensitive to the
employed motional model.\cite{Sholl_1981}
For long
times, ultimately, 
$G^\mathrm{n,HF}_\mathrm{inter}(t)$ exhibits a $t^{-3/2}$ scaling behavior
and can be expressed using the reduced time-scale $u$ as
\begin{equation}
\label{eq:Gu32}
\lim_{u\rightarrow \infty}G^\mathrm{n,HF}_\mathrm{inter}(u)\approx
\frac{1}{6\sqrt{\pi} \cdot u^{3/2}}\;.
\end{equation}
The accurate description of the $t^{-3/2}$ long-time limiting behavior of 
$G^\mathrm{n}_\mathrm{inter}(t)$ 
by means of MD simulations
is particularly 
important for properly describing 
the low frequency limit of the corresponding
spectral density function 
$\lim_{\omega\rightarrow 0} J^\mathrm{HF}_\mathrm{inter}(\omega)\propto \sqrt{\omega}$,
which can be utilised to extract the inter-diffusion coefficient
from the slope of frequency-dependent relaxation rate 
$\lim_{\omega\rightarrow 0} R_1(\omega)$ vs. 
$\sqrt{\omega}$ \cite{Hwang_1975}.
Using the normalized dipole-dipole correlation function
of the Hwang and Freed theory according to
Equation \ref{eq:g2model_u}, the 
full intermolecular spectral density 
of a random walker is given by 
\begin{eqnarray}
\hspace*{-20pt}
J^\mathrm{HF}_\mathrm{inter}(\omega_u)&=& \frac{2}{5}
\left<
  \sum_j r^{-6}_{ij}(0)
\right> 
\cdot 
J^\mathrm{n,HF}_\mathrm{inter}(\omega_u)\;.
\end{eqnarray}
Here $J^\mathrm{n,HF}_\mathrm{inter}(\omega_u)$ denotes
the ``normalized'' Hwang-Freed spectral density, obtained
as a Fourier transformation of $G^\mathrm{n}_\mathrm{inter}(t)$  with
\begin{eqnarray}
\hspace*{-20pt}
J^\mathrm{n,HF}_\mathrm{inter}(\omega_u)&=& 
\frac{54}{\pi}\cdot\frac{d^2}{D'}\int\limits_0^\infty 
\mathrm{Re}
\left\{
\int\limits_0^\infty 
e^{-u\cdot x^2}\; e^{i\omega_u u} 
\;
du 
\right\} 
\\ \nonumber 
&& \times \,\frac{x^2\;dx}{81+9x^2-2x^4+x^6} 
\end{eqnarray}
and
\begin{eqnarray}
\label{eq:HW_sdensity}
J^\mathrm{n,HF}_\mathrm{inter}(\omega_u)&=& 
\frac{54}{\pi}\cdot\frac{d^2}{D'}
\\ \nonumber
&& \times \int\limits_0^\infty 
\frac{dx}{(81+9x^2-2x^4+x^6)(1+\omega_u^2/x^4)}\;,
\end{eqnarray}
where $\omega_u\equiv\omega \cdot d^2/D'$ denotes a
reduced frequency scale, corresponding to the 
reduced time scale $u$.
From Equation \ref{eq:HW_sdensity} follows directly
the spectral density in the extreme narrowing limit 
as
\begin{eqnarray}
J^\mathrm{HF}_\mathrm{inter}(0) &=& \frac{2}{5}
\left<
  \sum_j r^{-6}_{ij}(0)
\right> 
\cdot \frac{4}{9}\cdot\frac{d^2}{D'}\;,
\end{eqnarray}
where 
\begin{equation}
\label{eq:taug}
\tau_\mathrm{G,HF}\!=\!J^\mathrm{n,HF}_\mathrm{inter}(0)\!=\!
\frac{4}{9}\cdot \frac{d^2}{D'}
\end{equation}
represents the 
intermolecular dipole-dipole
``correlation-time''
obtained as integral over the 
normalized dipole-dipole correlation function $G^\mathrm{n,HF}_\mathrm{inter}(t)$.
The limiting behavior 
of  the ``normalized'' spectral density
given by Equation \ref{eq:HW_sdensity}
for small frequencies is characterized by
a $\omega^{1/2}$ dependence according to
\begin{equation}
\label{eq:Jinter_sq}
\lim_{\omega_u\rightarrow 0}J^\mathrm{n,HF}_\mathrm{inter}(\omega_u)
\approx  J^\mathrm{n,HF}_\mathrm{inter}(0)
- \frac{\sqrt{2}}{6}\cdot \frac{d^2}{D'}\cdot \sqrt{\omega_u}\;.
\end{equation}

The Hwang and Freed theory outlined
above describes the behavior of ideal random walkers characterized
by infinitely long diffusion paths sampled from an infinitely large system.
In computer simulations of condensed matter systems, however, 
we mostly deal with finite system sizes
using periodic boundary conditions. These conditions impose 
the following two
problems: 1) they limit the volume from which the starting positions 
are sampled from, and, 2) the trajectories are altered by box-shifting, if
not unwrapped. Unwrapping the trajectories, however, has the
unfortunate tradeoff
of drastically reducing the accuracy of 
the computed short-time behavior \cite{Busch_2021} by
not allowing the particles to reconvene. Both problems can be countered
by increasing the system size, but they still might persist to some level. 
To thoroughly study and quantify both phenomena,
we show 
in \figurename\  \ref{fig:scalemaster}b the dipole-dipole correlation functions computed from Monte Carlo simulations with very short cutoff radii $R/d\leq4$. 
Note that both depicted correlation functions 
show a strong deviation from
the Hwang-Freed model for $t\rightarrow\infty$. 
This deviation is due a systematic depletion of particles
at long times $t$, and is related to  the lack of particles arriving from  starting distances with $r_{ij}(0)>R$, 
leading to an entirely different scaling behavior at long times.  
The computed $G^\mathrm{n}_\mathrm{inter}(u)$
scales with the square of ratio $(R/d)^2$ and approaches a $t^{-5/2}$ long-time limiting
behavior according to
\begin{equation}
\label{eq:g2cutofflong}
\lim_{u\rightarrow \infty}G^\mathrm{n}_\mathrm{inter}(u,R/d)\approx
\frac{1}{a\cdot u^{5/2}} \cdot \left(\frac{R}{d}\right)^2
\end{equation}
with $a\approx 23.6$. 
When PBCs are introduced, another effect comes into play: if a particle is leaving the box
and entering the box on the opposite side, it is basically changing its identity. 
If this identity change is ignored, the consequence is a change in 
the direction of the vector
connecting these two particles due to the 
mechanisms of the ``minimum image convention'' \cite{allentildesley}, 
which has obviously ramifications for the computed dipole-dipole correlation function $G^\mathrm{n}_\mathrm{inter}(u)$.
As can be seen in the green solid curve shown
 in \figurename\ \ref{fig:scalemaster}c, this 
mechanism even enhances the effect due to the
restricted sampling volume and leads to an even stronger deviation of the computed
$G^\mathrm{n}_\mathrm{inter}(u)$ from the $t^{-3/2}$-behavior. 
The example of $R/d\!=\!L/(2d)\!=\!4$ shown in \figurename\ \ref{fig:scalemaster}c
roughly corresponds to a rather small but not unrealistic system size
of about 128 water molecules, according to the parameters given in 
\tablename\ \ref{tab:md1} for $273\,\mbox{K}$.
From \figurename\ \ref{fig:scalemaster}c it is also evident, however,
that for sufficiently small times (such as $u\approx 1$), the
additional deviation according
to the periodic boundary conditions is practically negligible compared
to the effect due to the limited sampling volumes.

In the following, we would like to derive a procedure to
determine up to which time interval we can actually trust 
the computed intermolecular dipolar correlation functions despite
the presence of periodic boundary conditions and  limited sampling volumes.
To approximate
the effect caused to the limited sampling volumes
on the correlation function according to the Hwang and Freed model given
in Equation \ref{eq:g2model_u},
we use a nonzero lower boundary value 
$x_\mathrm{P}\!=\!\theta\cdot d/R$ with
$\theta\!\approx\!2.53$
for the
integral of Equation \ref{eq:g2model_u}, leading to
\begin{equation}
\label{eq:g2model_u_approx}
G^\mathrm{n,HF}_\mathrm{inter}(u,x_\mathrm{P})=
\frac{1}{A(x_\mathrm{P})}\;
\int\limits_{x_\mathrm{P}}^\infty \frac{x^2\cdot e^{-u\cdot x^2}\;dx}{81+9x^2-2x^4+x^6}\; ,
\end{equation}
realising that the variable $x$ is essentially representing an inverse distance. Here
the parameter $\theta$ has been determined empirically to provide the best agreement
with our random walker simulations for various values of $R/d$.
Note that the normalisation constant $A(x_\mathrm{P})$ needs
to be computed by numerical integration, except for $A(x_\mathrm{P}\!=\!0)=\pi/54$.
The deviation of the approximate expression given by Equation \ref{eq:g2model_u_approx} from
Equation \ref{eq:g2model_u} can then be quantified by
\begin{equation}
\label{eq:scaling}
s(u,x_\mathrm{P})=\frac{G^\mathrm{n,HF}_\mathrm{inter}(u,x_\mathrm{P}\!=\!0)}{G^\mathrm{n,HF}_\mathrm{inter}(u,x_\mathrm{P})}\;,
\end{equation}
where the denominator represents Equation \ref{eq:g2model_u}. As shown in
\figurename\ \ref{fig:scale_u}a, Equation \ref{eq:scaling} very well captures
the initial effect due to the limited sampling volumes over a broad
range of $R/d$ values that would correspond to liquid water simulations
ranging from 2000 to about 16000 water molecules in a cubic unit cell.
However, as shown in \figurename\ \ref{fig:scale_u}b, for longer times, 
correcting the Monte Carlo simulation data via $s(u,x_\mathrm{P})$ leads
to an overcorrection, suggesting that the introduction
of a certain time-limit $t_\mathrm{tr}$ is necessary, up to which the correction could
be meaningfully applied. Realising that the corresponding timescale
is governed by the ratio of the radius $R$ (or the half box size $L/2$) and the 
inter-diffusion coefficient $D'$, we use here
\begin{equation}
t_\mathrm{tr}=\frac{R^2}{8\pi D'}
=\frac{L^2}{32\pi D'}\,,
\end{equation}
which will consistently result in a time range where the scaling function
$s(u,x_\mathrm{P})\leq 1.1$, as
shown in \figurename\ \ref{fig:scale_u}a. Employing the definition of 
the reduced timescale, we get
\begin{equation}
u_\mathrm{tr}=\frac{1}{8\pi}\cdot\frac{R^2}{d^2}
=\frac{1}{32\pi}\cdot\frac{L^2}{d^2}
\,,
\end{equation}
indicating that the trusted time interval is just defined by the
ratio $R/d$ (or $L/(2d)$). As shown in \figurename\ \ref{fig:scale_u}c,
when considering times $t\leq t_\mathrm{tr}$
the deviation introduced  additionally due to the
effect of periodic boundary conditions
can be practically neglected.
\begin{figure}
        \includegraphics[width=0.3\textwidth]{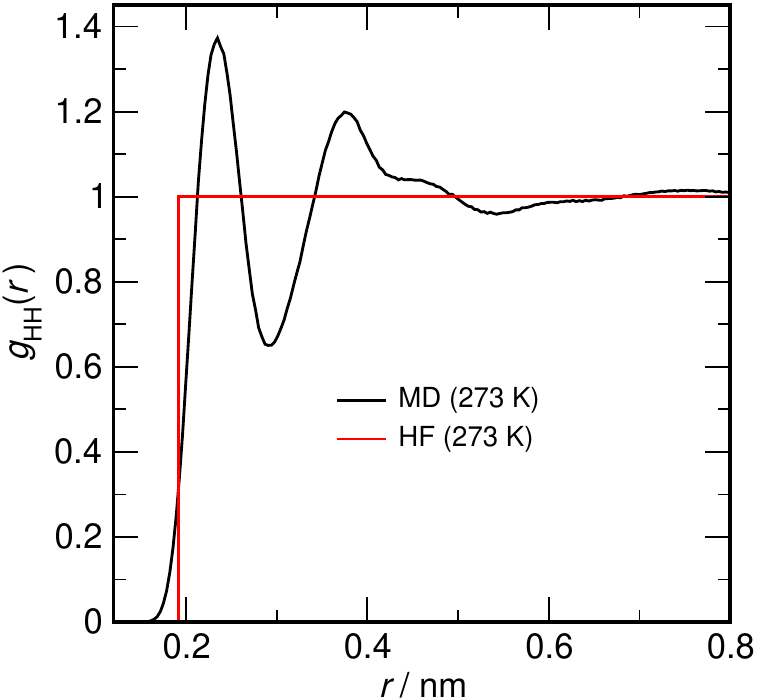}     
        \caption{\label{fig:gofr_hh}
	    Intermolecular H-H radial distribution function 
	    $g_\mathrm{HH}(r)$ 
	    as a function of distance $r$ for $273\,\mbox{K}$ 
	    in addition to the corresponding
	    step-like $g_\mathrm{HH}(r)$ according to
	    the Hwang and Freed theory with a 
	    distance of closest approach of $d_\mathrm{HH}\!=\!0.192\,\mbox{nm}$.}
\end{figure}
\begin{figure*}
        \includegraphics[width=0.305\textwidth]{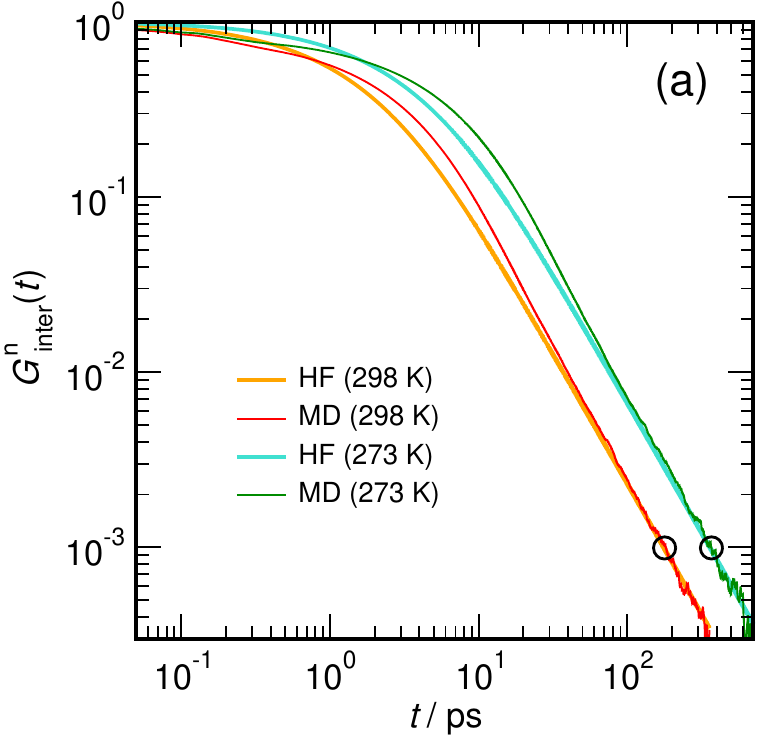}        
        \hspace*{1em}       
        \includegraphics[width=0.3\textwidth]{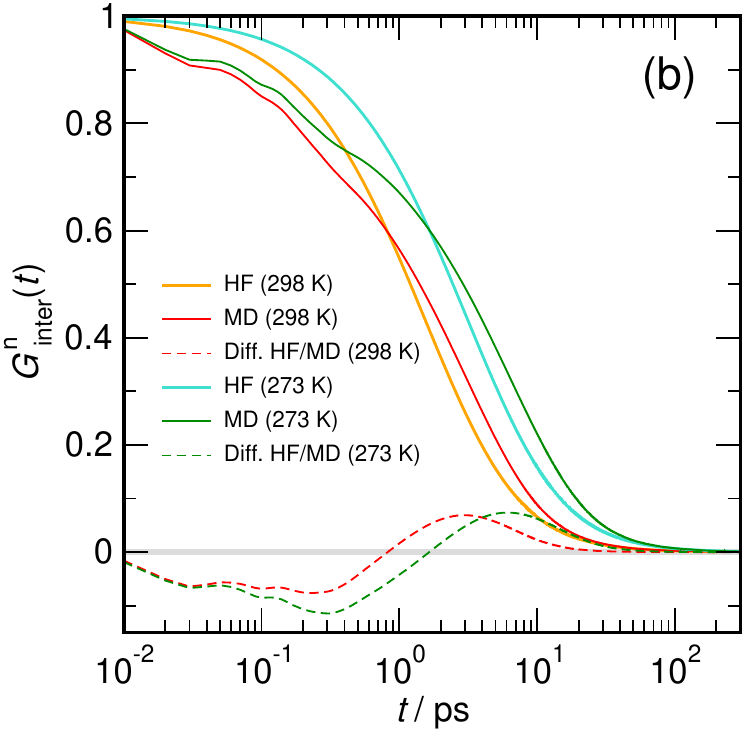}   
        \hspace*{1em}       
        \includegraphics[width=0.305\textwidth]{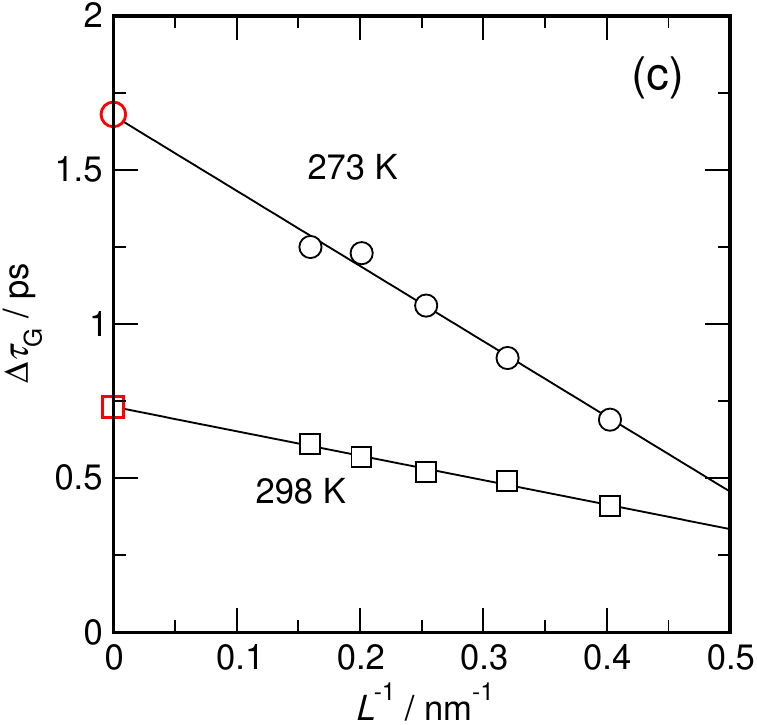}                     
        \caption{\label{fig:g2_water}
        Normalized intermolecular dipole-dipole correlation functions
        $G^\mathrm{n}_\mathrm{inter}(t)$ computed for TIP4P/2005 
        water for $273\,\mbox{K}$ and $298\,\mbox{K}$.       
        Red and green lines: $G^\mathrm{n,MD}_\mathrm{inter}
        (t)$ are obtained from MD simulations
        containing 8192 water molecules. Corrections according to Equation \ref{eq:scaling}
        are applied.
	    Orange and turquoise lines $G^\mathrm{n,HF}_\mathrm{inter}(t)$ computed using
	    data shown in Table \ref{tab:md1}.
	    Open circles: $t_\mathrm{tr}\!=\!L^2/(32\pi D')$
	    	    Dashed lines: difference functions 
	    $\Delta G^\mathrm{n}_\mathrm{inter}(t)=G^\mathrm{n,MD}_\mathrm{inter}(t)-G^\mathrm{n,HF}_\mathrm{inter}(t)$. 
	     a) Log-log plot. b) Linear-log plot
	     c) Scaling of the computed 
         intermolecular
         $\Delta \tau_\mathrm{G}$ 
         given in Table \ref{tab:md1}
         as a function of the inverse
         box length $L^{-1}$, analogous to the scaling of the translational diffusion
         coefficient suggested by Yeh and Hummer \cite{yeh_2004}.  
         Extrapolated values for $L\!\rightarrow\!\infty$ are indicated in red.
	    	    }
\end{figure*}
\begin{table*}
\caption{\label{tab:md1}
Parameters describing the intermolecular dipolar NMR relaxation
from MD simulations
under $NVT$ conditions at the indicated densities $\rho$ and temperatures $T$.
$L$: MD unit cell box-length.  
$d_\mathrm{HH}$: Distance of closest approach computed from H-H pair distribution
functions including a long-range correction.
$D_\text{E}$: ``Einstein'' water self-diffusion coefficient,
determined from the slope of the center-of-mass mean square displacement of the 
water molecules (slope fitted to time interval between $15\,\mbox{ps}$ and $200\,\mbox{ps}$).
$D_0$: water self-diffusion coefficient including the Yeh-Hummer finite-size 
correction \cite{yeh_2004} for systems with periodic boundary conditions
$D_0\!=\!D_\text{E}+k_\mathrm{B}T\zeta/(6\pi\eta L)$ with $\zeta\approx 2.837297$ and
$\eta$ the shear viscosity ($0.855\,\mbox{mPa\,s}$ at $298\,\mbox{K}$ and
$1.76\,\mbox{mPa\,s}$ at $273\,\mbox{K}$ \cite{Gonzales_2010}).
Inter-diffusion coefficient used for the correction: $D'\!=\!2 D_\text{E}$.
Maximum time interval up to which the computed 
and corrected dipolar correlation function can be trusted: 
$t_\mathrm{tr}\!=\!L^2/(32\pi D')$. 
Deviation of the total dipole-dipole correlation time from the Hwang-Freed model: 
$\Delta \tau_\text{G}$.
It is obtained by numerically integrating the
difference $\Delta G^\mathrm{n}_\mathrm{inter}(t)=G^\mathrm{n,MD}_\mathrm{inter}(t)-G^\mathrm{n,HF}_\mathrm{inter}(t)$ up to time $t_\mathrm{tr}$.
Total intermolecular dipole-dipole correlation time: 
$\tau_\text{G}\!=\!(4/9)\,d_\mathrm{HH}^2/(2D_0)+\Delta \tau_\text{G}$.
}
\setlength{\tabcolsep}{0.24cm}
        \centering              
\begin{tabular}{ccccccccccc}
\hline\hline\\[-0.2em]
$N$ & 
$T/\text{K}$ &
$\rho/\text{g}\,\text{cm}^{-3}$ &
$L/\text{nm}$ &
$d_\mathrm{HH}/\mathrm{\AA}$ &
$L/(2d)$ & 
$D_\text{E}/10^{-9}\,\text{m}^2\text{s}^{-1}$ & 
$D_0/10^{-9}\,\text{m}^2\text{s}^{-1}$ & 
$t_\mathrm{tr}/\mbox{ps}$&
$\Delta\tau_\mathrm{G}/\text{ps}$ &
$\tau_\text{G}/\text{ps}$ 
\\\hline\\[-0.6em]
512  & 273 & 0.9997 & 2.48368 & 1.92 &6.47& 0.98 & 1.11 & $62.6$ & 0.69 & 8.07 \\
1024 & 273 & 0.9997 & 3.12924 & 1.92 &8.15& 1.01 & 1.11 & $86.2$ & 0.89 & 8.27 \\
2048 & 273 & 0.9997 & 3.94259 & 1.92 &10.3& 1.03 & 1.11 & $150$  & 1.06 & 8.44 \\
4096 & 273 & 0.9997 & 4.96735 & 1.92 &12.9& 1.05 & 1.11 & $234$  & 1.23 & 8.61 \\
8192 & 273 & 0.9997 & 6.25847 & 1.92 &16.3& 1.06 & 1.11 & $368$  & 1.25 & 8.63\\
$\infty$ & 273 &  0.9997&  $\infty$  & 1.92 & $\infty$ & 1.11 & 1.11 & $\infty$ & 1.68 & 9.06\\[0.6em]
512  & 298 & 0.9972 & 2.48582 & 1.93 &6.44& 2.01 & 2.30 & $30.6$ & 0.41   & 4.01 \\
1024 & 298 & 0.9972 & 3.13194 & 1.93 &8.11& 2.08 & 2.31 & $46.9$ & 0.49   & 4.07 \\
2048 & 298 & 0.9972 & 3.94600 & 1.93 &10.2& 2.12 & 2.31 & $73.1$ & 0.52 & 4.10 \\
4096 & 298 & 0.9972 & 4.97165 & 1.93 &12.9& 2.16 & 2.31 & $114$  & 0.57 & 4.15 \\
8192 & 298 & 0.9972 & 6.26388 & 1.93 &16.2& 2.19 & 2.31 & $178$  & 0.61 & 4.19 \\
$\infty$ & 298 & 0.9972 & $\infty$ & 1.93 & $\infty$ & 2.31 & 2.31 & $\infty$ & 0.73 & 4.31 \\[0.6em]
\hline\hline
\end{tabular}
\end{table*}
\begin{figure*}
        \includegraphics[width=0.3\textwidth]{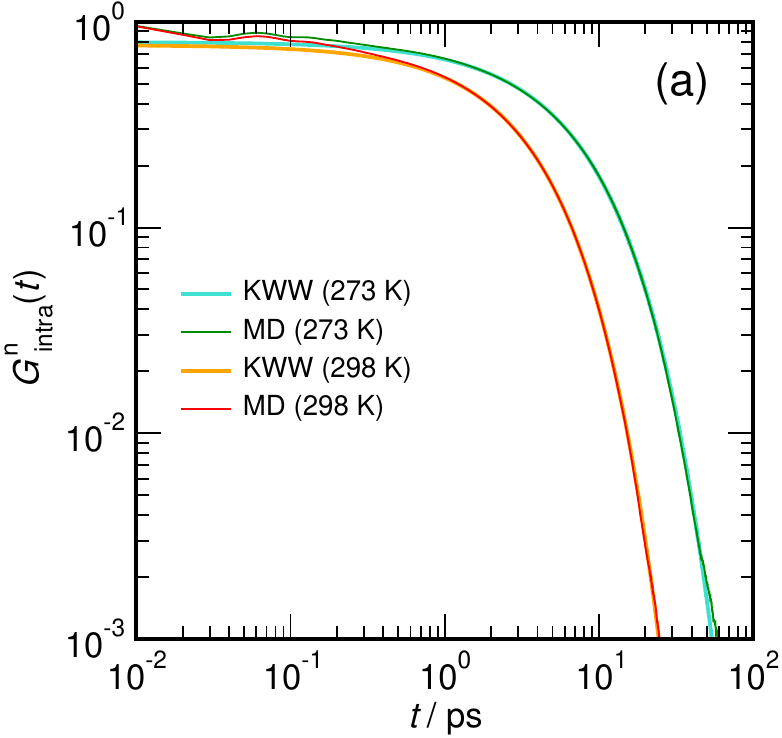}   
                \hspace*{2em}       	     
        \includegraphics[width=0.3\textwidth]{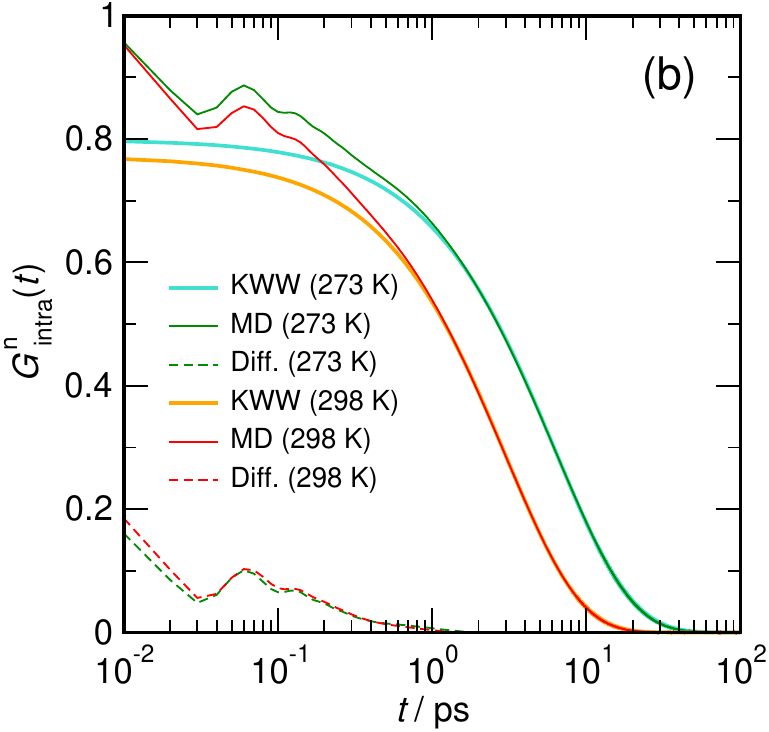}        
        \caption{\label{fig:g2_intra}
         Normalized intramolecular dipole-dipole correlation functions
        $G^\mathrm{n}_\mathrm{intra}(t)$ computed for TIP4P/2005 
        water for $273\,\mbox{K}$ and $298\,\mbox{K}$.   
        Red and green lines: $G^\mathrm{n,MD}_\mathrm{intra}(t)$ 
        are obtained from MD simulations
        containing 8192 water molecules. 
	    Orange and turquoise lines: KWW functions
	     $G^\mathrm{n,K}_\mathrm{intra}(t)$ 
	     according to Equation \ref{eq:KWW}
	     computed using
	    data shown in \tablename\ \ref{tab:md2}.
	    a) Log-log plot. b) Linear-log plot including
	    difference functions (dashed lines) 
	    $\Delta G^\mathrm{n}_\mathrm{intra}(t)=G^\mathrm{n,MD}_\mathrm{intra}(t)-G^\mathrm{n,K}_\mathrm{intra}(t)$.
	    }
\end{figure*}
\begin{table*}
\caption{\label{tab:md2}
Parameters describing the intramolecular dipolar NMR relaxation
from MD simulations.
Due to the fixed intramolecular distance $r_\mathrm{HH}\!=\!1.514\,\mbox{\AA}$
in the TIP4P/2005 water model,
the computed intramolecular dipole-dipole correlation functions
are equivalent to the reorientational correlation functions of the
intramolecular H--H vector. The computed $G^\mathrm{n,MD}_\mathrm{intra}(t)$ are fitted
to a Kohlrausch-Williams-Watts (KWW) function with
$G^\mathrm{n,K}_\mathrm{intra}(t)\!=\!A_\mathrm{K}\cdot\exp[-(t/\tau_\mathrm{K})^{\beta_\mathrm{K}}]$.
The fit is performed for data within a time window between
$1\,\mbox{ps}$ and $100\,\mbox{ps}$.
The integrated correlation time of the KWW function is
given as 
$\tau_\mathrm{G,K}\!=\!A_\mathrm{K}\tau_\mathrm{K}\beta^{-1}_\mathrm{K}\Gamma(\beta^{-1}_\mathrm{K})$,
where $\Gamma(\ldots)$ represents the Gamma-function.
Deviation of the total dipole-dipole correlation time from the KWW function: 
$\Delta \tau_\text{G}$.
It is obtained by numerically integrating the
difference $\Delta G^\mathrm{n}_\mathrm{intra}(t)=G^\mathrm{n,MD}_\mathrm{intra}(t)-G^\mathrm{n,K}_\mathrm{intra}(t)$ up to
a time of $5\,\mbox{ps}$.
Total intramolecular dipole-dipole correlation time: $\tau_\text{G}\!=\tau_\mathrm{G,K}+\Delta \tau_\text{G}$.
}
\setlength{\tabcolsep}{0.71cm}
        \centering              
\begin{tabular}{cccccccc}
\hline\hline\\[-0.2em]
$N$ & 
$T/\text{K}$ &
$A_\mathrm{K}$ &
$\tau_\mathrm{K}/\mathrm{ps}$ &
$\beta_\mathrm{K}$ &
$\tau_\mathrm{G,K}/\mathrm{ps}$ & 
$\Delta \tau_\mathrm{G}/\mbox{ps}$ &
$\tau_\text{G}/\mbox{ps}$ 
\\\hline\\[-0.6em]
512  & 273 & 0.800  & 6.12 & 0.886 & 5.20 & 0.024 & 5.22  \\
1024 & 273 & 0.811  & 6.12 & 0.866 & 5.34 & 0.021 & 5.35  \\
2048 & 273 & 0.795  & 6.32 & 0.892 & 5.31 & 0.028 & 5.34  \\
4096 & 273 & 0.798  & 6.31 & 0.886 & 5.34 & 0.028 & 5.37  \\
8192 & 273 & 0.799  & 6.30 & 0.884 & 5.35 & 0.028 & 5.38  \\[0.6em]
512  & 298 & 0.779  & 3.06 & 0.908 & 2.50 & 0.006 & 2.51  \\
1024 & 298 & 0.767  & 3.07 & 0.916 & 2.45 & 0.029 & 2.48  \\
2048 & 298 & 0.771  & 3.02 & 0.908 & 2.44 & 0.031 & 2.47  \\
4096 & 298 & 0.778  & 2.99 & 0.899 & 2.45 & 0.025 & 2.48  \\
8192 & 298 & 0.772  & 3.03 & 0.907 & 2.45 & 0.028 & 2.48  \\[0.6em]\hline\hline
\end{tabular}
\end{table*}
\begin{figure*}
        \includegraphics[width=0.3\textwidth]{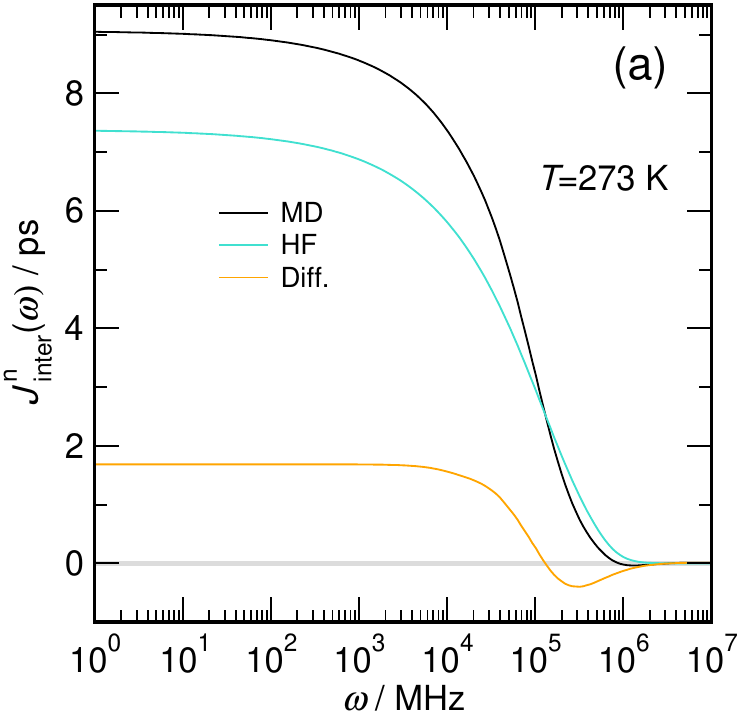}     
                \hspace*{0.5em}          
        \includegraphics[width=0.3\textwidth]{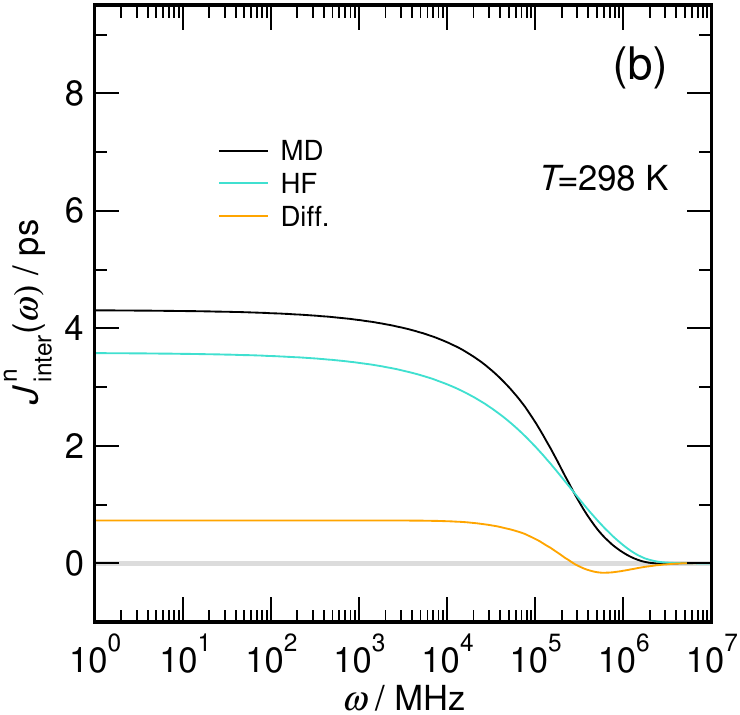}   
                \hspace*{0.5em}          
        \includegraphics[width=0.3\textwidth]{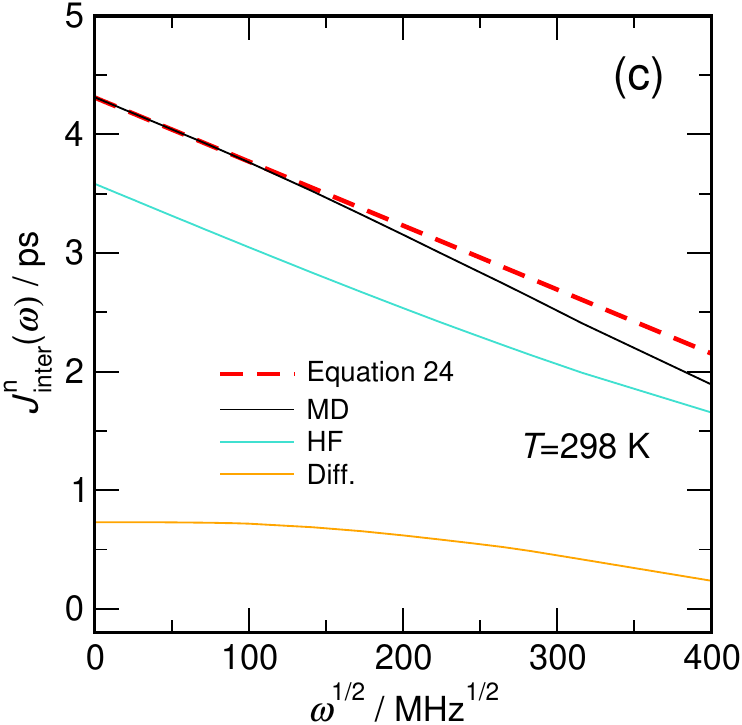}                  
        \caption{\label{fig:Jomega_inter}
         Intermolecular spectral density
        $J^\mathrm{n}_\mathrm{inter}(\omega)$ computed for TIP4P/2005 
        water for a) $273\,\mbox{K}$ and b) $298\,\mbox{K}$.   
        Black lines: $J^\mathrm{n,MD}_\mathrm{inter}(\omega)$.
        Turquoise lines: $J^\mathrm{n,HF}_\mathrm{inter}(\omega)$,
	     computed using
	    data shown in Table 
        \ref{tab:md1} for $N\rightarrow\infty$. 
	    Orange lines: difference functions 
	    $\Delta J^\mathrm{n}_\mathrm{inter}(\omega)=
	    J^\mathrm{n,MD}_\mathrm{inter}(\omega)-J^\mathrm{n,HF}_\mathrm{inter}(\omega)$,
		obtained by Fourier transformation 
		of $\Delta G^\mathrm{n}_\mathrm{inter}(t)$
		from MD simulations
        containing 8192 water molecules.
        c) Intermolecular spectral density
        obtained for $298\,\mbox{K}$
        as a function of
        $\sqrt{\omega}$.
        Red dashed line: Low-frequency limiting behavior 
        $\lim_{\omega\rightarrow 0}J^\mathrm{n,HF}_\mathrm{inter}(\omega)$
        according to Equation \ref{eq:Jinter_sq}.
        }
\end{figure*}
\begin{figure*}
        \includegraphics[width=0.3\textwidth]{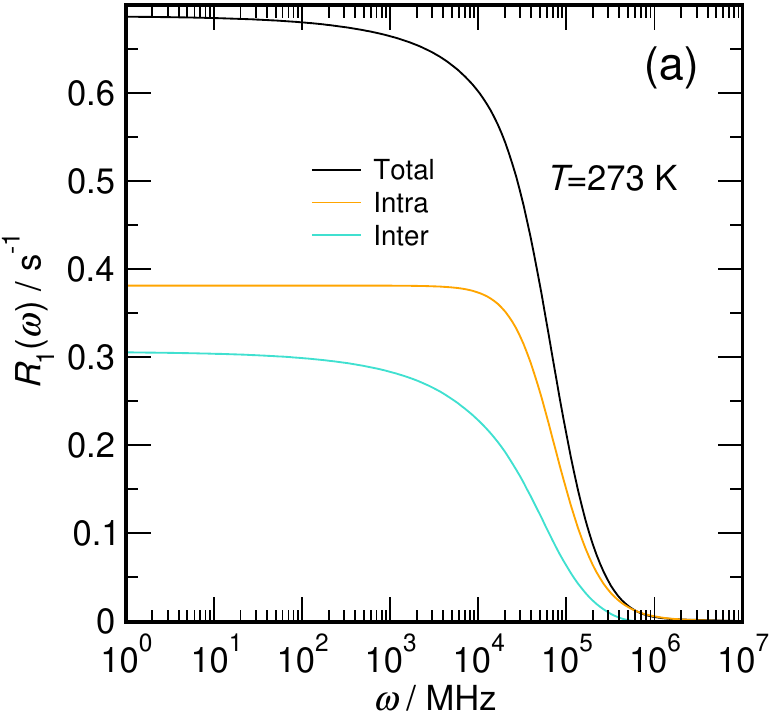}     
                \hspace*{2em}          
        \includegraphics[width=0.3\textwidth]{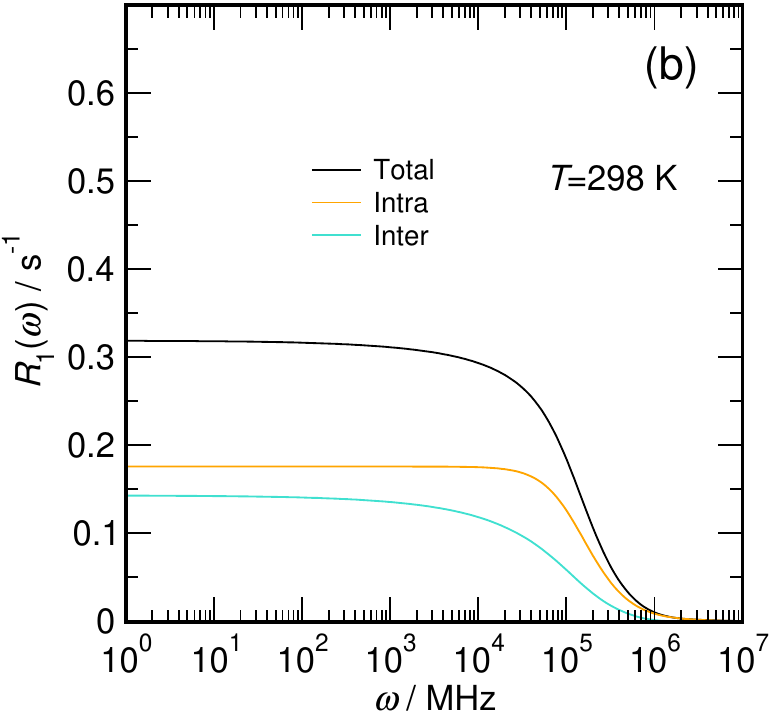}        
        \caption{\label{fig:Jomega_all}
        Frequency-dependent
        intermolecular, intramolecular, and total 
        dipole-dipole $^1$H NMR 
        relaxation rates 
        computed for TIP4P/2005 
        water
        according to Equations \ref{eq:R1_inter} and
        \ref{eq:R1_intra}
         for a) $273\,\mbox{K}$ and b) $298\,\mbox{K}$.}  
\end{figure*}

\begin{table}
\caption{\label{tab:md3}
Intermolecular, Intramolecular, and total
dipolar $^1$H NMR relaxation rates 
in the extreme narrowing limit
computed for
the TIP4P/2005 water model.
Intramolecular H-H distance of the TIP4P/2005 model used for computing the
intramolecular relaxation rate: $r_\mathrm{HH}\!=\!1.514\,\mbox{\AA}$.
Experimental relaxation rates were obtained by Krynicki \cite{Krynicki_1966}.
Goldammer and Zeidler \cite{Goldammer_1969}
studied mixtures of water with organic compounds to separate intra- 
from intermolecular contributions.
Calero et al.\cite{Calero_2015} computed inter- and intramolecular
relaxation rates from MD simulations of the TIP4P/2005 water model.
}
\setlength{\tabcolsep}{0.06cm}
        \centering              
\begin{tabular}{ccccc}
\hline\hline\\[-0.2em]
$N$ & 
$T/\text{K}$ &
$R_{1,\mathrm{inter}}(0)/\mbox{s}^{-1}$ &
$R_{1,\mathrm{intra}}(0)/\mbox{s}^{-1}$ &
$R_{1}(0)/\mbox{s}^{-1}$ 
\\\hline\\[-0.6em]
512  & 273 & 0.272   & 0.370 & 0.642      \\
1024 & 273 & 0.279   & 0.380 & 0.659      \\
2048 & 273 & 0.285   & 0.379 & 0.664      \\
4096 & 273 & 0.291   & 0.381 & 0.672   \\
8192 & 273 & 0.292   & 0.382 & 0.674      \\
$\infty$   & 273 & 0.306   &  0.382     &  0.688      \\
Expt.\cite{Krynicki_1966} & 273 & --  & -- & 0.578 \\[0.6em]
512  & 298 & 0.133     & 0.178 & 0.311      \\
1024 & 298 & 0.135     & 0.176 & 0.311      \\
2048 & 298 & 0.136     & 0.175 & 0.311      \\
4096 & 298 & 0.138     & 0.176 & 0.314   \\
8192 & 298 & 0.139     & 0.176 & 0.315      \\ 
$\infty$   & 298 & 0.143   &  0.176    & 0.319        \\[0.6em]
Expt.\cite{Goldammer_1969,Krynicki_1966} & 298 & 0.110  & 0.170 & 0.280   \\
MD\cite{Calero_2015}    & 298 & 0.087  & 0.176 & 0.263 \\[0.6em]\hline\hline
\end{tabular}
\end{table}
\begin{table}
\caption{\label{tab:md4}
Intermolecular, Intramolecular, and total
dipolar $^1$H NMR relaxation rates as a function of the
frequency $\omega$
computed for
the TIP4P/2005 water model.}
\setlength{\tabcolsep}{0.12cm}
        \centering              
\begin{tabular}{ccccc}
\hline\hline\\[-0.2em]
$\omega/\mbox{MHz}$ & 
$T/\text{K}$ &
$R_{1,\mathrm{inter}}(\omega)/\mbox{s}^{-1}$ &
$R_{1,\mathrm{intra}}(\omega)/\mbox{s}^{-1}$ &
$R_{1}(\omega)/\mbox{s}^{-1}$ 
\\\hline\\[-0.6em]
0    & 273 & 0.306  & 0.382 & 0.688      \\
50   & 273 & 0.301  & 0.382 & 0.683      \\
200  & 273 & 0.296  & 0.382 & 0.678    \\
400  & 273 & 0.292  & 0.382 & 0.674      \\
800  & 273 & 0.286  & 0.382 & 0.668   \\
1200 & 273 & 0.281  & 0.382 & 0.663    \\[0.6em]
0    & 298 & 0.143  & 0.176 & 0.319      \\
50   & 298 & 0.141  & 0.176 & 0.317      \\
200  & 298 & 0.140  & 0.176 & 0.316      \\
400  & 298 & 0.138  & 0.176 & 0.314   \\
800  & 298 & 0.136  & 0.176 & 0.312   \\
1200 & 298 & 0.135  & 0.176 & 0.311    \\[0.6em]\hline\hline
\end{tabular}
\end{table}

\subsection{Using a Mixed Theory/MD Approach to Compute the
Frequency-Dependent NMR Relaxation of $^1$H Nuclei in Liquid Water}

We have performed $NVT$ MD simulations of TIP4P/2005 water at 
$273\,\mbox{K}$ and $298\,\mbox{K}$
 at respective densities of 
$0.9997\,\text{g}\,\text{cm}^{-3}$  and
$0.9972\,\text{g}\,\text{cm}^{-3}$, 
corresponding to an average pressure of about $1\,\mbox{bar}$ for
system sizes between 512 and 8192 molecules.
Data characterizing the simulations can be found in 
\tablename\ \ref{tab:md1}. We have computed the self-diffusion coefficients using
the Einstein formula  \cite{allentildesley} according to
\begin{equation}
\label{eq:De}
D_\textrm{E}
=
\frac{1}{6}
\frac{\partial}{\partial t}
\lim_{t\rightarrow\infty}
\left<
|\mathbf{r}(0)
-
\mathbf{r}(t)
|^2
\right>\;,
\end{equation}
where $\mathbf{r}(t)=[r_x(t),r_y(t),r_z(t)]$ represents the position of the center of mass
of a water molecule at time $t$.
All computed self-diffusion coefficients 
shown \tablename\ \ref{tab:md1} were
determined from the slope of the mean square
displacement of the water molecules
fitted to time intervals between $15\,\mbox{ps}$ and $200\,\mbox{ps}$.
Note that the $D_\textrm{E}$ is a system size dependent quantity,
which can, however, be corrected via \cite{duenweg_1993,yeh_2004}
\begin{equation}
\label{eq:Ds}
D_0 = D_\text{E}+\frac{k_\mathrm{B}T\zeta}{6\pi\eta L}\;,
\end{equation}
with the box size $L$, and the shear viscosity $\eta$.
Here, $D_0$ is the
system size independent {\em true} self-diffusion coefficient obtained
for $L\rightarrow\infty$,
$k_\mathrm{B}$ represents Boltzmann's constant and $T$ is the temperature.
The parameter $\zeta\!\approx\!2.837297$ 
is the analogue to a Madelung constant 
\cite{beenacker_1986}
of a cubic lattice, which can be
computed via Ewald summation.\cite{beenacker_1986,hasimoto_1959} 
All computed values for $D_0$ are also given in \tablename\ \ref{tab:md1}.
To perform the correction, we have employed the
shear viscosity $\eta$ of $0.855\,\mbox{mPa\,s}$ at $298\,\mbox{K}$ and
$1.76\,\mbox{mPa\,s}$ at $273\,\mbox{K}$ reported by
Ref. \cite{Gonzales_2010}. The distances of closest approach 
$d_\mathrm{HH}$
for 
the $^1$H nuclei given in \tablename\ \ref{tab:md1}
for $273\,\mbox{K}$ and $289\,\mbox{K}$
are determined by integrating the $r^{-6}$ weighted
H-H radial distribution functions according to Equations \ref{eq:dhh-1}
and \ref{eq:dhh-2}. The numerical integration of the
pair correlation function was performed up to
a distance of $L/2$ and was improved by adding
a term for the long-range correction of $32\pi/(3L^3)$.
Note the slight temperature dependence of the computed $d_\mathrm{HH}$.
Both the radial distribution function $g_\mathrm{HH}(r)$ obtained from MD
and according to the Hwang Freed theory are shown in \figurename\ \ref{fig:gofr_hh}
for $273\,\mbox{K}$.

To determine the intermolecular dipolar relaxation
correlation functions $G^\mathrm{n,MD}_\mathrm{inter}(t)$, we have 
computed an average over
$512\times(N-1)$ intermolecular correlation functions
where $N$ is the number of molecules, leading to a total
of 4193792 correlation functions for the 8192 molecule system. 
For the calculation, we have used one H-atom per water molecule.
In \figurename\ \ref{fig:g2_water}a we are comparing the 
time dependence of the normalized
intermolecular dipole-dipole correlation functions $G^\mathrm{n,MD}_\mathrm{inter}(t)$
computed directly from molecular simulations including the correction
according to Equation \ref{eq:scaling} with the prediction of the Hwang
and Freed model $G^\mathrm{n,HF}_\mathrm{inter}(t)$ employing the
distances of closest approach $d_\mathrm{HH}$ and
the inter-diffusion coefficients $D'=2D_\mathrm{E}$ obtained
for a system size of 8192 water molecules. The values for the
computed trusted time-intervals
$t_\mathrm{tr}$ are indicated by open circles and are also given 
in \tablename\ \ref{tab:md1} for all system sizes and temperatures.
For times $t\approx t_\mathrm{tr}$
the function shows a $t^{-3/2}$ scaling behavior and the curve determined
from MD simulation asymptotically approaches the Hwang and Freed model.
For times $t\geq t_\mathrm{tr}$ both curves are practically indistinguishable.
A log-linear representation of the data, including the
difference function
\begin{equation}
\Delta G^\mathrm{n}_\mathrm{inter}(t)=G^\mathrm{n,MD}_\mathrm{inter}(t)-G^\mathrm{n,HF}_\mathrm{inter}(t)
\end{equation}
is shown in \figurename\ \ref{fig:g2_water}b. Note that the difference
function is negative up to a time of about $1\,\mbox{ps}$, then turns positive until 
it asymptotically approaches zero. The negative region is due to fast
librational motions of the water molecules, whereas the  positive region
is related due to a resting tendency of the protons after  large angular jumps. 
In total, both
the negative and positive deviation from an overall continuous diffusion of the
$^1$H nuclei, as described by the Hwang-Freed model,
are reflecting the jump-like reorientational dynamics of
water molecules discussed in detail by D. Laage and J.T. Hynes.\cite{laage_2006,laage_2008}
The intermolecular 
correlation time $\tau_\mathrm{G}$ can be computed 
as an integral over $G^\mathrm{n,MD}_\mathrm{inter}(t)$, which can be splitted into two terms
according to
\begin{equation}
\tau_\mathrm{G} = \tau_\mathrm{G,HF} + \Delta\tau_\mathrm{G}
\end{equation}
with $\tau_\mathrm{G,HF}=9/4\cdot d_\mathrm{HH}/D'$
following Equation \ref{eq:taug}. Here $\Delta\tau_\mathrm{G}$
can be computed comfortably via numerical integration of
\begin{equation}
\Delta\tau_\mathrm{G} \approx \int\limits_0^{t_\mathrm{tr}} 
\Delta G^\mathrm{n}_\mathrm{inter}(t) \;dt
\end{equation}
due to the short-time nature of $\Delta G^\mathrm{n}_\mathrm{inter}(t)$.
Computed values for $\Delta\tau_\mathrm{G}$ 
and $\tau_\mathrm{G}$ 
are listed in \tablename\ \ref{tab:md1}
for all temperatures and system sizes.
Note that the inter-diffusion coefficients used for determining
$\tau_\mathrm{G,HF}$ are based here on 
the system size dependent diffusion coefficients $D'=2D_\mathrm{E}$ shown in
\tablename\ \ref{tab:md1}. This is a necessary requirement, since otherwise 
$\Delta G^\mathrm{n,MD}_\mathrm{inter}(t)$ 
and 
$\Delta G^\mathrm{n,HF}_\mathrm{inter}(t)$
would not match at long times.
As a consequence, $\Delta\tau_\mathrm{G}$ shows a system size dependence,
as it is indicated in \figurename\ \ref{fig:g2_water}c. Here, the apparent
linear dependence from the inverse box length is purely based on empirical evidence.
The rationale for an increase of $\Delta\tau_\mathrm{G}$ is  based on the fact that
the initial decay  of $G^\mathrm{n,MD}_\mathrm{inter}(t)$ is largely due to the mutual reorientational
motions of adjacent molecules and that
that dynamics of
these reorientational motions is nearly system size independent \cite{celebi_2021},
thus increasing the net-positive difference between
$G^\mathrm{n,MD}_\mathrm{inter}(t)$ and $G^\mathrm{n,HF}_\mathrm{inter}(t)$ with increasing system size. Based
on the apparent linear $L^{-1}$-dependence, we can also give an estimate for
$\Delta\tau_\mathrm{G}$ for $L\rightarrow\infty$. In combination with
the true self-diffusion coefficient $D_0$, we can give an estimate for the
{\em true} system size independent correlation time 
$\tau_\mathrm{G}$ shown in \tablename\ \ref{tab:md1}, and can thus
also give an estimate for the {\em true} intermolecular relaxation rate
$\lim_{L\rightarrow\infty}R_\mathrm{inter}(0)$.

To describe the intramolecular dipolar relaxation, we 
essentially compute the reorientational motion of
the H-H vector, since the H-H distance of $r_\mathrm{HH}=0.1514\,\mbox{nm}$ is 
fixed within the TIP4P/2005 water model. Here the computed
$G^\mathrm{n,MD}_\mathrm{intra}(t)$ represent averages over all $N$ intramolecular
H-H vectors. In principle, we 
choose to
follow the same strategy
for the intramolecular dipolar correlation as we did for the 
intermolecular dynamics.
The main difference, however, is that we 
do not employ a physics-based mechanistic model, but choose to
apply the empricial Kohlrausch-Williams Watts (KWW)
function for describing the long time behavior
\begin{equation}
\label{eq:KWW}
G^\mathrm{n,K}_\mathrm{intra}(t) = A_\mathrm{K}\cdot\exp\left[-  \left(\frac{t}{\tau_\mathrm{K}}\right)^{\beta_\mathrm{K}}\right]\;.
\end{equation}
This empirical model  is fitted to the computed $G^\mathrm{n,MD}_\mathrm{intra}(t)$
over a time interval between $1\,\mbox{ps}$  and $100\,\mbox{ps}$.
In \figurename\ \ref{fig:g2_intra}a both functions are plotted and they
become pretty much indistinguishable for times $t$ larger than about $2\,\mbox{ps}$.
The fitted parameters are summarized in 
\tablename\ \ref{tab:md2}.
A log-linear representation of the data, including the
difference function
\begin{equation}
\Delta G^\mathrm{n}_\mathrm{intra}(t)=G^\mathrm{n,MD}_\mathrm{intra}(t)-G^\mathrm{n,K}_\mathrm{intra}(t)
\end{equation}
is shown in \figurename\ \ref{fig:g2_intra}b.
Significant differences between 
$G^\mathrm{n,MD}_\mathrm{intra}(t)$ and $G^\mathrm{n,K}_\mathrm{intra}(t)$ are restricted to
a time-interval $t\leq 1 \,\mbox{ps}$.
Hence
the intramolecular
correlation time $\tau_\mathrm{G}$ was computed 
as an integral over $G^\mathrm{n,MD}_\mathrm{intra}(t)$, 
which can be splitted into two terms
according to
\begin{equation}
\label{eq:deltagintra}
\tau_\mathrm{G} = \tau_\mathrm{G,K} + \Delta\tau_\mathrm{G}
\end{equation}
with 
\begin{equation}
\label{eq:tauKWW}
\tau_\mathrm{G,K}\!=\!A_\mathrm{K}\tau_\mathrm{K}\beta^{-1}_\mathrm{K}\Gamma(\beta^{-1}_\mathrm{K})\;,
\end{equation}
where $\Gamma(\ldots)$ represents the Gamma-function.
The deviation of the total dipole-dipole correlation time from the KWW function 
$\Delta \tau_\text{G}$
is obtained by numerically integrating 
the difference $\Delta G^\mathrm{n}_\mathrm{intra}(t)$ 
according to Equation \ref{eq:deltagintra}
up to a time of $5\,\mbox{ps}$.
Both the fitted parameters and the computed total correlation time
$\tau_\mathrm{G}$ shown in \tablename\ \ref{tab:md2} 
obtained for various system sizes
do not indicate any
system size dependence, which is in accordance with the 
finding of Celebi et al.\ \cite{celebi_2021} 
who noticed
that the finite size correction
for the rotational diffusion scales with the inverse box volume and is
therefore much smaller than the one for translational diffusion.

Next, we want to compute the frequency-dependent spectral densities and
thus the frequency-dependent relaxation rates. To compute the intermolecular
spectral density from MD simulation, we use
\begin{equation}
J^\mathrm{n,MD}_\mathrm{inter}(\omega)
=
J^\mathrm{n,HF}_\mathrm{inter}(\omega)
+
\Delta J^\mathrm{n}_\mathrm{inter}(\omega)
\end{equation}
with
\begin{equation}
\label{eq:deltaj_md}
\Delta J^\mathrm{n}_\mathrm{inter}(\omega)
\approx
\int\limits_0^{t_\mathrm{tr}} 
\Delta G^\mathrm{n}_\mathrm{inter}(\omega)
\cos (\omega t) \;dt\;.
\end{equation}
Here the integration 
in Equation \ref{eq:deltaj_md}
is performed numerically 
employing the trapezoidal rule
up the time $t_\mathrm{tr}$, where
both functions $G^\mathrm{n,MD}_\mathrm{inter}(\omega)$
and $G^\mathrm{n,HF}_\mathrm{inter}(\omega)$ are deemed indistinguishable.
An important feature of this approach is that arbitrary 
frequencies $\omega$ can be used
here, which is helpful in evaluating the relaxation rate, where 
both $J^\mathrm{n,MD}_\mathrm{inter}(\omega)$ and 
$J^\mathrm{n,MD}_\mathrm{inter}(2\omega)$ need to be computed.
To properly predict $J^\mathrm{n,MD}_\mathrm{inter}(\omega)$ for systems
with $L\rightarrow\infty$, we employ the system size independent
self-diffusion coefficient $D_0$ for computing $J^\mathrm{n,HF}_\mathrm{inter}(\omega)$.
In addition, we use
\begin{equation}
\Delta G^\mathrm{n}_\mathrm{inter}(\omega)
\approx
\frac{\tau_{\mathrm{G},N\rightarrow\infty}}{\tau_{\mathrm{G},N=8192}}
\cdot
\Delta G^\mathrm{n}_{\mathrm{inter},N=8192}(\omega)
\end{equation}
to predict the behavior
of $\Delta G^\mathrm{n}_\mathrm{inter}(\omega)$ for infinite system sizes.
Here $\Delta G^\mathrm{n}_{\mathrm{inter},N=8192}(\omega)$ is the 
difference function computed for a system containing 8192 water molecules, and
$\tau_{\mathrm{G},N=8192}=J^\mathrm{n,MD}_{\mathrm{inter},N=8192}(0)$ 
and $\tau_{\mathrm{G},N\rightarrow\infty}=J^\mathrm{n,MD}_{\mathrm{inter},N\rightarrow\infty}(0)$ are the
corresponding correlation times predicted for an infinite system size via
extrapolation shown in \tablename\ \ref{tab:md1}. The frequency depdendence
of 
$J^\mathrm{n,MD}_\mathrm{inter}(\omega)$,
$J^\mathrm{n,HF}_\mathrm{inter}(\omega)$,
and
$\Delta J^\mathrm{n}_\mathrm{inter}(\omega)$ are shown
in \figurename\ \ref{fig:Jomega_inter} for
$273\,\mbox{K}$ and $298\,\mbox{K}$.
Note that the low frequency-behavior of the intermolecular
spectral density 
$\lim_{\omega\rightarrow 0}J^\mathrm{n,HF}_\mathrm{inter}(\omega)$
shown in \figurename\ \ref{fig:Jomega_inter}c
follows a $\sqrt{\omega}$ dependence
according to Equation \ref{eq:Jinter_sq}
up to about $100\,\mbox{MHz}^{1/2}$, which corresponds to
frequencies up to about $10\,\mbox{GHz}$, which is way beyond the 
frequency 
range accessible via currently available NMR technology.
We can therefore conclude that frequency-dependent
$^1$H NMR relaxation observed experimentally for
liquid water at $298\,\mbox{K}$
is largely
dominated by translational diffusion.
Note, however, that the dispersion of the computed
$J^\mathrm{n,MD}_\mathrm{inter}(\omega)$ shown in 
\figurename\ \ref{fig:Jomega_inter}a and 
\figurename\ \ref{fig:Jomega_inter}b is markedly
deviating from the behavior predicted by Hwang and Freed,
showing a much sharper decay. This is the consequence
of the negative part of the 
$\Delta J^\mathrm{n}_\mathrm{inter}(\omega)$-functions
observed for frequencies $\omega\geq 1.5\times 10^5\,\mbox{MHz}$ 
and $\omega\geq 3\times 10^5\,\mbox{MHz}$ observed for 
$T\!=\!273\,\mbox{K}$ and
$T\!=\!298\,\mbox{K}$, respectively.
This spectral feature is obviously
related to fast
librational motions of the water molecules, in combination with  
large angular jumps, which
are characterizing their reorientational motions.

Next, we would apply the same strategy
outlined in the previous paragraph
 also to the intramolecular
relaxation rate, and then
combine both intra- and intermolecular contributions
to describe the total  $^1$H relaxation.
To compute the intramolecular
spectral density from MD simulation, we use
\begin{equation}
J^\mathrm{n,MD}_\mathrm{intra}(\omega)
=
J^\mathrm{n,K}_\mathrm{intra}(\omega)
+
\Delta J^\mathrm{n}_\mathrm{intra}(\omega)
\end{equation}
with
\begin{equation}
\label{eq:deltaj_md_reor}
\Delta J^\mathrm{n}_\mathrm{intra}(\omega)
\approx
\int\limits_0^{t^*} 
\Delta G^\mathrm{n}_\mathrm{intra}(\omega)
\cos (\omega t) \;dt\;.
\end{equation}
Here the computation of the integral
in Equation \ref{eq:deltaj_md_reor}
is as well performed numerically,
employing the trapezoidal rule
up a time $t^*=5\mbox{ps}$, where
both functions $G^\mathrm{n,MD}_\mathrm{intra}(\omega)$
and $G^\mathrm{n,K}_\mathrm{intra}(\omega)$ become effectively indistinguishable.
The Fourier-transform of $G^\mathrm{n,K}_\mathrm{intra}(t)$, defined in
Equation \ref{eq:KWW}, $J^\mathrm{n,K}_\mathrm{intra}(\omega)$, needs, however, to be computed numerically, 
due to the lack of an analytical Fourier-transform
equivalent of the KWW-function. 
To compute $J^\mathrm{n,K}_\mathrm{intra}(\omega)$ properly,
we have tested the convergence 
of the numerical cosine-transform evaluation by comparing it to the
limiting value for $\omega\rightarrow 0$ provided by the analytically
obtained data given in \tablename\ \ref{tab:md2}.
In \figurename\ \ref{fig:Jomega_all} we have plotted 
the inter- and intramolecular contribution to the
$^1$H NMR relaxation rate
following Equation \ref{eq:relax}
computed via
\begin{eqnarray}
\label{eq:R1_inter}
R_{1,\mathrm{inter}}(\omega)
&=&
\gamma_\mathrm{H}^4\hbar^2 \cdot\frac{3}{4}\cdot\left(\frac{\mu_0}{4\pi}\right)^2\cdot
\frac{8\pi}{15}\cdot\frac{\rho_\mathrm{H}}{d_\mathrm{HH}^3}\times\\
&&\left\{
J^\mathrm{n,MD}_\mathrm{inter}(\omega)
+
4\,J^\mathrm{n,MD}_\mathrm{inter}(2\omega)
\right\}
\nonumber
\end{eqnarray}
and
\begin{eqnarray}
\label{eq:R1_intra}
R_{1,\mathrm{intra}}(\omega)
&=&
\gamma_\mathrm{H}^4\hbar^2 \cdot\frac{3}{4}\cdot\left(\frac{\mu_0}{4\pi}\right)^2\cdot
\frac{2}{5}\cdot
\frac{1}{r_\mathrm{HH}^6}\times\\
&&\left\{
J^\mathrm{n,MD}_\mathrm{intra}(\omega)
+
4\,J^\mathrm{n,MD}_\mathrm{intra}(2\omega)
\right\}
\nonumber
\end{eqnarray}
as a function of the frequency $\omega$ over many orders of magnitude. 
Here $\rho_\mathrm{H}$ is representing the number density of
the $^1$H nuclei in liquid water. Note
that for frequencies $\omega\leq 10^3\,\mbox{MHz}$
$R_{1,\mathrm{intra}}(\omega)$ is basically frequency independent, confirming
that the dispersion that can be experimentally obtained is solely caused
by the intermolecular contributions.
In \tablename\ \ref{tab:md3} we have listed the data for the
inter-, intramolecular, and total
dipolar $^1$H NMR relaxation rates 
in the extreme narrowing limit $\omega\rightarrow 0$
computed for
the TIP4P/2005 water model. Note that the computed
relaxation rates are significantly larger than the 
experimental data. This might seem odd since the diffusion coefficient
of the TIP4P/2005 model at $298\,\mbox{K}$
of $D_0=2.31\times 10^{-9}\,\mbox{m}^2\,\mbox{s}^{-1}$ 
matches almost perfectly the experimental value of
$D_0=2.3\times 10^{-9}\,\mbox{m}^2\,\mbox{s}^{-1}$.\cite{krynicki_1978}
However, one has to keep in mind that the TIP4P/2005 model is
a rigid model, and that additional high frequency
bond-length and bond-bending motions, which the TIP4P/2005 model is lacking, will
lead to a further quenching of both 
$G^\mathrm{n,MD}_\mathrm{inter}(t)$ and
$G^\mathrm{n,MD}_\mathrm{intra}(t)$, leading to smaller inter- and intramolecular
relaxation rates. Therefore it should not come as a surprise that the 
computed $^1$H relaxation rates are larger than the experimental ones.
It is, however, surprising that the $^1$H relaxation rate computed by
Calero et al.\ \cite{Calero_2015} for the TIP4P/2005 model is actually smaller than
the experimental data of Krynicky  \cite{Krynicki_1966}.
By using the data given in their paper, we have computed their
intramolecular relaxation rate 
and found it to be in perfect agreement with our data (see \tablename\ \ref{tab:md3}).
Their intermolecular relaxation rate of 
$0.087\,\mbox{s}^{-1}$, however, is even smaller than the
value according to Hwang-Freed theory purely based on intermolecular
diffusion of $0.119\;\mbox{s}^{-1}$ when using  
$D_0=2.31\times 10^{-9}\,\mbox{m}^2\,\mbox{s}^{-1}$ and 
$d_\mathrm{HH}=0.193\,\mbox{nm}$. This, however, seems rather unlikely and
emphasizes the importance to properly consider the long-time
nature of the $G^\mathrm{n,MD}_\mathrm{inter}(t)$ with its slowly
decaying $t^{-3/2}$ time dependence.

Finally, in \tablename\ \ref{tab:md4} we report data for the computed
relaxation rates for TIP4P/2005 water for frequencies accessible via
modern NMR hardware. The reduction of the total relaxation rate
of $3.6\%$ at $273\,\mbox{K}$
and 
$2.5\%$ at $298\,\mbox{K}$ is purely due to changes in intermolecular
relaxation rate. Of course, for supercooled liquid water, the effect of
dispersion will increase substantially. Note that
Krynicky  determined the $^1$H NMR relaxation rate of water
at a frequency of $50\,\mbox{MHz}$.\cite{Krynicki_1966} The expected
reduction of the relaxation rate due to intermolecular contributions
of $0.7\%$ 
compared to the true extreme narrowing limit
is, however, smaller than their reported experimental error
of about $1\%$.

\section{Conclusion}

We have introduced a computational framework aimed at accurately 
determining the frequency-dependent 
intermolecular NMR dipole-dipole relaxation 
rate of spin $1/2$ nuclei through MD simulations. 
This framework circumvents the influence of well-known 
finite-size effects on translational diffusion. 
Moreover, we have developed a method to manage and 
rectify the impacts stemming from fixed distance-sampling 
cutoffs and periodic boundary conditions.

Our approach is capable of accurately 
forecasting the 
proper low-frequency $\sqrt{\omega}$-scaling behavior
of the intermolecular NMR dipole-dipole relaxation rate
observed experimentally.
It is based on the theory
of Hwang and Freed \cite{Hwang_1975}
for the 
intermolecular
dipole-dipole relaxation and is utilizing
their analytical expressions for both 
the dipole-dipole
correlation function and its corresponding
spectral density.
Deviations from the Hwang and Freed theory
caused by periodic boundary conditions and restrictions 
due to sampling distance cutoffs 
were studied and quantified by means of random walker
Monte Carlo simulations. These simulation were 
designed to perfectly replicate the force free hard sphere model 
underlying the Hwang and Freed theory.
Based on both the Hwang and Freed theory and the Monte
Carlo simulations,
an expression has been derived for
correcting for those effects and to determine
the time interval up to which the corrected correlation
functions faithfully follow the true
behavior observed, when restrictions 
due to sampling distance cutoffs  and periodic boundary effects
are absent.

As a proof of principle, our approach is demonstrated by computing the
frequency-dependent inter- and intramolecular dipolar NMR
relaxation rate of the
$^1$H nuclei in liquid water 
at $273\,\mbox{K}$ and $298\,\mbox{K}$
based on
simulations of
the TIP4P/2005 model. In particular, our calculations suggest that 
the intermolecular contribution to the $^1$H relaxation 
rate of the TIP4P/2005 model 
in the extreme narrowing limit
has been previously significantly 
underestimated.

\section*{Acknowledgements}

AS acknowledges funding by the Deutsche Forschungsgemeinschaft (DFG),
Project-No.\ 459405854.
The authors would like to thank the computer center at the University of 
Rostock (ITMZ) for providing and maintaining computational resources. 
 
\section*{Author Declarations}
 
\subsection*{Conflict of Interest}

The authors have no conflicts to disclose.
 
\subsection*{Author Contributions}

\textbf{Dietmar Paschek}: Conceptualization (lead); 
Methodology (lead); 
Formal analysis (lead); 
Investigation (lead); 
Software (lead);
Data curation (lead);
Visualization (lead); 
Supervision (equal);
Project administration (supporting);
Writing -- original draft (lead); 
Writing -- review \& editing (lead).
\textbf{Johanna Busch}: 
Investigation (supporting); 
Software (supporting);
Writing -- review \& editing (supporting). 
\textbf{Eduard Mock}: 
Software (supporting).
\textbf{Ralf Ludwig}: 
Conceptualization (supporting); 
Resources (lead);
Writing -- review \& editing (supporting).
\textbf{Anne Strate}: 
Conceptualization (supporting); 
Funding acquisition (lead); 
Project administration (lead); 
Supervision (equal);
Writing -- review \& editing (supporting). 

\section*{Data Availability}

The codes  of 
\href{https://www.gromacs.org}{GROMACS} 
and
\href{https://github.com/Paschek-Lab/MDorado}{MDorado}
are freely available. 
Input parameter and topology files for the MD simulations, the 
code for performing random walker Monte Carlo Simulations,
and the code for computing 
intermolecular 
and intramolecular
relaxation rates 
can be downloaded from GitHub via
\href{https://github.com/Paschek-Lab/DrelaxD}{github.com/Paschek-Lab/DrelaxD}.

\bibliography{all}

\begin{thebibliography}{48}%
\makeatletter
\providecommand \@ifxundefined [1]{%
 \@ifx{#1\undefined}
}%
\providecommand \@ifnum [1]{%
 \ifnum #1\expandafter \@firstoftwo
 \else \expandafter \@secondoftwo
 \fi
}%
\providecommand \@ifx [1]{%
 \ifx #1\expandafter \@firstoftwo
 \else \expandafter \@secondoftwo
 \fi
}%
\providecommand \natexlab [1]{#1}%
\providecommand \enquote  [1]{``#1''}%
\providecommand \bibnamefont  [1]{#1}%
\providecommand \bibfnamefont [1]{#1}%
\providecommand \citenamefont [1]{#1}%
\providecommand \href@noop [0]{\@secondoftwo}%
\providecommand \href [0]{\begingroup \@sanitize@url \@href}%
\providecommand \@href[1]{\@@startlink{#1}\@@href}%
\providecommand \@@href[1]{\endgroup#1\@@endlink}%
\providecommand \@sanitize@url [0]{\catcode `\\12\catcode `\$12\catcode
  `\&12\catcode `\#12\catcode `\^12\catcode `\_12\catcode `\%12\relax}%
\providecommand \@@startlink[1]{}%
\providecommand \@@endlink[0]{}%
\providecommand \url  [0]{\begingroup\@sanitize@url \@url }%
\providecommand \@url [1]{\endgroup\@href {#1}{\urlprefix }}%
\providecommand \urlprefix  [0]{URL }%
\providecommand \Eprint [0]{\href }%
\providecommand \doibase [0]{http://dx.doi.org/}%
\providecommand \selectlanguage [0]{\@gobble}%
\providecommand \bibinfo  [0]{\@secondoftwo}%
\providecommand \bibfield  [0]{\@secondoftwo}%
\providecommand \translation [1]{[#1]}%
\providecommand \BibitemOpen [0]{}%
\providecommand \bibitemStop [0]{}%
\providecommand \bibitemNoStop [0]{.\EOS\space}%
\providecommand \EOS [0]{\spacefactor3000\relax}%
\providecommand \BibitemShut  [1]{\csname bibitem#1\endcsname}%
\let\auto@bib@innerbib\@empty
\bibitem [{\citenamefont {Abragam}(1961)}]{Abragam1961}%
  \BibitemOpen
  \bibfield  {author} {\bibinfo {author} {\bibfnamefont {A.}~\bibnamefont
  {Abragam}},\ }\href@noop {} {\emph {\bibinfo {title} {The Principles of
  Nuclear Magnetism}}}\ (\bibinfo  {publisher} {Oxford University Press},\
  \bibinfo {year} {1961})\BibitemShut {NoStop}%
\bibitem [{\citenamefont {Kowalewski}(2013)}]{kowalewski_2013}%
  \BibitemOpen
  \bibfield  {author} {\bibinfo {author} {\bibfnamefont {J.}~\bibnamefont
  {Kowalewski}},\ }\enquote {\bibinfo {title} {Nuclear magnetic resonance},}\ \
  (\bibinfo  {publisher} {Royal Society of Chemistry},\ \bibinfo {year}
  {2013})\ Chap.\ \bibinfo {chapter} {Nuclear spin relaxation in liquids and
  gases}, pp.\ \bibinfo {pages} {230--275}\BibitemShut {NoStop}%
\bibitem [{\citenamefont {Kruk}, \citenamefont {Meier},\ and\ \citenamefont
  {R{\"o}ssler}(2011)}]{Kruk_2011}%
  \BibitemOpen
  \bibfield  {author} {\bibinfo {author} {\bibfnamefont {D.}~\bibnamefont
  {Kruk}}, \bibinfo {author} {\bibfnamefont {R.}~\bibnamefont {Meier}}, \ and\
  \bibinfo {author} {\bibfnamefont {A.}~\bibnamefont {R{\"o}ssler}},\
  }\bibfield  {title} {\enquote {\bibinfo {title} {Translational and rotational
  diffusion of glycerol by means of field cycling $^1${H} {NMR} relaxometry},}\
  }\href@noop {} {\bibfield  {journal} {\bibinfo  {journal} {J. Phys. Chem. B}\
  }\textbf {\bibinfo {volume} {115}},\ \bibinfo {pages} {951--957} (\bibinfo
  {year} {2011})}\BibitemShut {NoStop}%
\bibitem [{\citenamefont {Kruk}, \citenamefont {Hermann},\ and\ \citenamefont
  {R{\"o}ssler}(2012)}]{Kruk_2012}%
  \BibitemOpen
  \bibfield  {author} {\bibinfo {author} {\bibfnamefont {D.}~\bibnamefont
  {Kruk}}, \bibinfo {author} {\bibfnamefont {A.}~\bibnamefont {Hermann}}, \
  and\ \bibinfo {author} {\bibfnamefont {E.~A.}\ \bibnamefont {R{\"o}ssler}},\
  }\bibfield  {title} {\enquote {\bibinfo {title} {Field-cycling {NMR}
  relaxometry of viscous liquids and polymers},}\ }\href@noop {} {\bibfield
  {journal} {\bibinfo  {journal} {Prog. Nucl. Magn. Reson. Spectrosc.}\
  }\textbf {\bibinfo {volume} {63}},\ \bibinfo {pages} {33--64} (\bibinfo
  {year} {2012})}\BibitemShut {NoStop}%
\bibitem [{\citenamefont {Sholl}(1981)}]{Sholl_1981}%
  \BibitemOpen
  \bibfield  {author} {\bibinfo {author} {\bibfnamefont {C.~A.}\ \bibnamefont
  {Sholl}},\ }\bibfield  {title} {\enquote {\bibinfo {title} {Nuclear-spin
  relaxation by translational diffusion in liquids and solids - high-frequency
  and low-frequency limits},}\ }\href@noop {} {\bibfield  {journal} {\bibinfo
  {journal} {J. Phys. C: Solid State Phys.}\ }\textbf {\bibinfo {volume}
  {14}},\ \bibinfo {pages} {447--464} (\bibinfo {year} {1981})}\BibitemShut
  {NoStop}%
\bibitem [{\citenamefont {Hwang}\ and\ \citenamefont
  {Freed}(1975)}]{Hwang_1975}%
  \BibitemOpen
  \bibfield  {author} {\bibinfo {author} {\bibfnamefont {L.-P.}\ \bibnamefont
  {Hwang}}\ and\ \bibinfo {author} {\bibfnamefont {J.~H.}\ \bibnamefont
  {Freed}},\ }\bibfield  {title} {\enquote {\bibinfo {title} {Dynamic effects
  of pair correlation functions on spin relaxation by translational diffusion
  in liquids},}\ }\href@noop {} {\bibfield  {journal} {\bibinfo  {journal} {J.
  Chem. Phys.}\ }\textbf {\bibinfo {volume} {63}},\ \bibinfo {pages}
  {4017--4025} (\bibinfo {year} {1975})}\BibitemShut {NoStop}%
\bibitem [{\citenamefont {Overbeck}\ \emph {et~al.}(2020)\citenamefont
  {Overbeck}, \citenamefont {Golub}, \citenamefont {Schr{\"o}der},
  \citenamefont {Appelhagen}, \citenamefont {Paschek}, \citenamefont
  {Neymeyr},\ and\ \citenamefont {Ludwig}}]{overbeck_2020}%
  \BibitemOpen
  \bibfield  {author} {\bibinfo {author} {\bibfnamefont {V.}~\bibnamefont
  {Overbeck}}, \bibinfo {author} {\bibfnamefont {B.}~\bibnamefont {Golub}},
  \bibinfo {author} {\bibfnamefont {H.}~\bibnamefont {Schr{\"o}der}}, \bibinfo
  {author} {\bibfnamefont {A.}~\bibnamefont {Appelhagen}}, \bibinfo {author}
  {\bibfnamefont {D.}~\bibnamefont {Paschek}}, \bibinfo {author} {\bibfnamefont
  {K.}~\bibnamefont {Neymeyr}}, \ and\ \bibinfo {author} {\bibfnamefont
  {R.}~\bibnamefont {Ludwig}},\ }\bibfield  {title} {\enquote {\bibinfo {title}
  {Probing relaxation models by means of fast field-cycling relaxometry, {NMR}
  spectroscopy and molecular dynamics simulations: Detailed insight into the
  translational and rotational dynamics of a protic ionic liquid},}\
  }\href@noop {} {\bibfield  {journal} {\bibinfo  {journal} {J. Mol. Liq.}\
  }\textbf {\bibinfo {volume} {319}},\ \bibinfo {pages} {114207} (\bibinfo
  {year} {2020})}\BibitemShut {NoStop}%
\bibitem [{\citenamefont {Overbeck}\ \emph {et~al.}(2021)\citenamefont
  {Overbeck}, \citenamefont {Appelhagen}, \citenamefont {R{\"o}{\ss}ler},
  \citenamefont {Niemann},\ and\ \citenamefont {Ludwig}}]{overbeck_2021}%
  \BibitemOpen
  \bibfield  {author} {\bibinfo {author} {\bibfnamefont {V.}~\bibnamefont
  {Overbeck}}, \bibinfo {author} {\bibfnamefont {A.}~\bibnamefont
  {Appelhagen}}, \bibinfo {author} {\bibfnamefont {R.}~\bibnamefont
  {R{\"o}{\ss}ler}}, \bibinfo {author} {\bibfnamefont {T.}~\bibnamefont
  {Niemann}}, \ and\ \bibinfo {author} {\bibfnamefont {R.}~\bibnamefont
  {Ludwig}},\ }\bibfield  {title} {\enquote {\bibinfo {title} {Rotational
  correlation times, diffusion coefficients and quadrupolar peaks of the protic
  ionic liquid ethylammonium nitrate by means of 1{H} fast field cycling {NMR}
  relaxometry},}\ }\href@noop {} {\bibfield  {journal} {\bibinfo  {journal} {J.
  Mol. Liq.}\ }\textbf {\bibinfo {volume} {322}},\ \bibinfo {pages} {114983}
  (\bibinfo {year} {2021})}\BibitemShut {NoStop}%
\bibitem [{\citenamefont {Westlund}\ and\ \citenamefont
  {{Lynden-Bell}}(1987)}]{Westlund:1987}%
  \BibitemOpen
  \bibfield  {author} {\bibinfo {author} {\bibfnamefont {P.}~\bibnamefont
  {Westlund}}\ and\ \bibinfo {author} {\bibfnamefont {R.}~\bibnamefont
  {{Lynden-Bell}}},\ }\bibfield  {title} {\enquote {\bibinfo {title} {A
  molecular dynamics study of the intermolecular spin-spin dipole-dipole
  correlation function of liquid acetonitrile},}\ }\href@noop {} {\bibfield
  {journal} {\bibinfo  {journal} {J. Magn. Reson.}\ }\textbf {\bibinfo {volume}
  {72}},\ \bibinfo {pages} {522--531} (\bibinfo {year} {1987})}\BibitemShut
  {NoStop}%
\bibitem [{\citenamefont {Schnitker}\ and\ \citenamefont
  {Geiger}(1987)}]{schnitker_1987}%
  \BibitemOpen
  \bibfield  {author} {\bibinfo {author} {\bibfnamefont {J.}~\bibnamefont
  {Schnitker}}\ and\ \bibinfo {author} {\bibfnamefont {A.}~\bibnamefont
  {Geiger}},\ }\bibfield  {title} {\enquote {\bibinfo {title} {{NMR}-quadrupole
  relaxation of {Xenon}-131 in water. a molecular dynamics simulation study},}\
  }\href@noop {} {\bibfield  {journal} {\bibinfo  {journal} {Z. Phys. Chem.}\
  }\textbf {\bibinfo {volume} {155}},\ \bibinfo {pages} {29--54} (\bibinfo
  {year} {1987})}\BibitemShut {NoStop}%
\bibitem [{\citenamefont {Yeh}\ and\ \citenamefont {Hummer}(2004)}]{yeh_2004}%
  \BibitemOpen
  \bibfield  {author} {\bibinfo {author} {\bibfnamefont {I.-C.}\ \bibnamefont
  {Yeh}}\ and\ \bibinfo {author} {\bibfnamefont {G.}~\bibnamefont {Hummer}},\
  }\bibfield  {title} {\enquote {\bibinfo {title} {System-size dependence of
  diffusion coefficients and viscosities from molecular dynamics simulations
  with periodic boundary conditions},}\ }\href@noop {} {\bibfield  {journal}
  {\bibinfo  {journal} {J. Phys. Chem. B}\ }\textbf {\bibinfo {volume} {108}},\
  \bibinfo {pages} {15873--15879} (\bibinfo {year} {2004})}\BibitemShut
  {NoStop}%
\bibitem [{\citenamefont {Moultos}\ \emph {et~al.}(2016)\citenamefont
  {Moultos}, \citenamefont {Zhang}, \citenamefont {Tsimpanogiannis},
  \citenamefont {Economou},\ and\ \citenamefont {Maginn}}]{moultos_2016}%
  \BibitemOpen
  \bibfield  {author} {\bibinfo {author} {\bibfnamefont {O.~A.}\ \bibnamefont
  {Moultos}}, \bibinfo {author} {\bibfnamefont {Y.}~\bibnamefont {Zhang}},
  \bibinfo {author} {\bibfnamefont {I.~O.}\ \bibnamefont {Tsimpanogiannis}},
  \bibinfo {author} {\bibfnamefont {I.~G.}\ \bibnamefont {Economou}}, \ and\
  \bibinfo {author} {\bibfnamefont {E.~J.}\ \bibnamefont {Maginn}},\ }\bibfield
   {title} {\enquote {\bibinfo {title} {System-size corrections for
  self-diffusion coefficients calculated from molecular dynamics simulations:
  The case of {CO}$_2$, n-alkanes, and poly(ethylene glycol) dimethyl
  ethers},}\ }\href@noop {} {\bibfield  {journal} {\bibinfo  {journal} {J.
  Chem. Phys.}\ }\textbf {\bibinfo {volume} {145}},\ \bibinfo {pages} {074109}
  (\bibinfo {year} {2016})}\BibitemShut {NoStop}%
\bibitem [{\citenamefont {Busch}\ and\ \citenamefont
  {Paschek}(2023)}]{busch_2023c}%
  \BibitemOpen
  \bibfield  {author} {\bibinfo {author} {\bibfnamefont {J.}~\bibnamefont
  {Busch}}\ and\ \bibinfo {author} {\bibfnamefont {D.}~\bibnamefont
  {Paschek}},\ }\bibfield  {title} {\enquote {\bibinfo {title} {{OrthoBoXY}: A
  simple way to compute true self-diffusion coefficients from {MD} simulations
  with periodic boundary conditions without prior knowledge of the
  viscosity},}\ }\href@noop {} {\bibfield  {journal} {\bibinfo  {journal} {J.
  Phys. Chem. B}\ }\textbf {\bibinfo {volume} {127}},\ \bibinfo {pages}
  {7983--7987} (\bibinfo {year} {2023})}\BibitemShut {NoStop}%
\bibitem [{\citenamefont {Honegger}\ \emph {et~al.}(2021)\citenamefont
  {Honegger}, \citenamefont {{Di Pietro}}, \citenamefont {Castiglione},
  \citenamefont {Vaccarini}, \citenamefont {Quant}, \citenamefont
  {Steinhauser}, \citenamefont {Schr\"oder},\ and\ \citenamefont
  {Mele}}]{honegger_2021}%
  \BibitemOpen
  \bibfield  {author} {\bibinfo {author} {\bibfnamefont {P.}~\bibnamefont
  {Honegger}}, \bibinfo {author} {\bibfnamefont {M.~E.}\ \bibnamefont {{Di
  Pietro}}}, \bibinfo {author} {\bibfnamefont {F.}~\bibnamefont {Castiglione}},
  \bibinfo {author} {\bibfnamefont {C.}~\bibnamefont {Vaccarini}}, \bibinfo
  {author} {\bibfnamefont {A.}~\bibnamefont {Quant}}, \bibinfo {author}
  {\bibfnamefont {O.}~\bibnamefont {Steinhauser}}, \bibinfo {author}
  {\bibfnamefont {C.}~\bibnamefont {Schr\"oder}}, \ and\ \bibinfo {author}
  {\bibfnamefont {A.}~\bibnamefont {Mele}},\ }\bibfield  {title} {\enquote
  {\bibinfo {title} {The intermolecular {NOE} depends on isotope selection:
  Short range vs. long range behavior},}\ }\href@noop {} {\bibfield  {journal}
  {\bibinfo  {journal} {J. Phys. Chem. Lett.}\ }\textbf {\bibinfo {volume}
  {12}},\ \bibinfo {pages} {8658--8663} (\bibinfo {year} {2021})}\BibitemShut
  {NoStop}%
\bibitem [{\citenamefont {Abascal}\ and\ \citenamefont
  {Vega}(2005)}]{abascal_2005}%
  \BibitemOpen
  \bibfield  {author} {\bibinfo {author} {\bibfnamefont {J.~L.~F.}\
  \bibnamefont {Abascal}}\ and\ \bibinfo {author} {\bibfnamefont
  {C.}~\bibnamefont {Vega}},\ }\bibfield  {title} {\enquote {\bibinfo {title}
  {A general purpose model for the condensed phases of water: {TIP4P/2005}},}\
  }\href@noop {} {\bibfield  {journal} {\bibinfo  {journal} {J. Chem. Phys.}\
  }\textbf {\bibinfo {volume} {123}},\ \bibinfo {pages} {234505} (\bibinfo
  {year} {2005})}\BibitemShut {NoStop}%
\bibitem [{\citenamefont {Calero}, \citenamefont {Mart{\'i}},\ and\
  \citenamefont {Gu{\`a}rdia}(2015)}]{Calero_2015}%
  \BibitemOpen
  \bibfield  {author} {\bibinfo {author} {\bibfnamefont {C.}~\bibnamefont
  {Calero}}, \bibinfo {author} {\bibfnamefont {J.}~\bibnamefont {Mart{\'i}}}, \
  and\ \bibinfo {author} {\bibfnamefont {E.}~\bibnamefont {Gu{\`a}rdia}},\
  }\bibfield  {title} {\enquote {\bibinfo {title} {$^1${H} nuclear spin
  relaxation of liquid water from molecular dynamics simulations},}\
  }\href@noop {} {\bibfield  {journal} {\bibinfo  {journal} {J. Phys. Chem. B}\
  }\textbf {\bibinfo {volume} {119}},\ \bibinfo {pages} {1966--1973} (\bibinfo
  {year} {2015})}\BibitemShut {NoStop}%
\bibitem [{\citenamefont {Odelius}\ \emph {et~al.}(1993)\citenamefont
  {Odelius}, \citenamefont {Laaksonen}, \citenamefont {Levitt},\ and\
  \citenamefont {Kowalewski}}]{Odelius:1993}%
  \BibitemOpen
  \bibfield  {author} {\bibinfo {author} {\bibfnamefont {M.}~\bibnamefont
  {Odelius}}, \bibinfo {author} {\bibfnamefont {A.}~\bibnamefont {Laaksonen}},
  \bibinfo {author} {\bibfnamefont {M.~H.}\ \bibnamefont {Levitt}}, \ and\
  \bibinfo {author} {\bibfnamefont {J.}~\bibnamefont {Kowalewski}},\ }\bibfield
   {title} {\enquote {\bibinfo {title} {Intermolecular dipole-dipole
  relaxation. a molecular dynamics simulation},}\ }\href@noop {} {\bibfield
  {journal} {\bibinfo  {journal} {J. Magn. Reson. Ser. A}\ }\textbf {\bibinfo
  {volume} {105}},\ \bibinfo {pages} {289--294} (\bibinfo {year}
  {1993})}\BibitemShut {NoStop}%
\bibitem [{\citenamefont {Hertz}(1967)}]{Hertz1967}%
  \BibitemOpen
  \bibfield  {author} {\bibinfo {author} {\bibfnamefont {H.~G.}\ \bibnamefont
  {Hertz}},\ }\bibfield  {title} {\enquote {\bibinfo {title} {Microdynamic
  {B}ehaviour of {L}iquids as studied by {NMR} {R}elaxation {T}imes},}\
  }\href@noop {} {\bibfield  {journal} {\bibinfo  {journal} {Prog. NMR Spec.}\
  }\textbf {\bibinfo {volume} {3}},\ \bibinfo {pages} {159--230} (\bibinfo
  {year} {1967})}\BibitemShut {NoStop}%
\bibitem [{\citenamefont {Hertz}\ and\ \citenamefont
  {Tutsch}(1976)}]{Hertz1976}%
  \BibitemOpen
  \bibfield  {author} {\bibinfo {author} {\bibfnamefont {H.~G.}\ \bibnamefont
  {Hertz}}\ and\ \bibinfo {author} {\bibfnamefont {R.}~\bibnamefont {Tutsch}},\
  }\bibfield  {title} {\enquote {\bibinfo {title} {Model {O}rientation
  {D}ependent {P}air {D}istribution {F}unctions {D}escribing the {A}ssociation
  of {S}imple {C}aboxyle {A}cids and of {E}thanol in {A}queous {S}olution},}\
  }\href@noop {} {\bibfield  {journal} {\bibinfo  {journal} {Ber. Bunsenges.
  Phys. Chem.}\ }\textbf {\bibinfo {volume} {80}},\ \bibinfo {pages}
  {1268--1278} (\bibinfo {year} {1976})}\BibitemShut {NoStop}%
\bibitem [{\citenamefont {Stejksal}\ \emph {et~al.}(1959)\citenamefont
  {Stejksal}, \citenamefont {Woessner}, \citenamefont {Farrar},\ and\
  \citenamefont {Gutowsky}}]{Stejksal:1959}%
  \BibitemOpen
  \bibfield  {author} {\bibinfo {author} {\bibfnamefont {E.~O.}\ \bibnamefont
  {Stejksal}}, \bibinfo {author} {\bibfnamefont {D.~E.}\ \bibnamefont
  {Woessner}}, \bibinfo {author} {\bibfnamefont {T.~C.}\ \bibnamefont
  {Farrar}}, \ and\ \bibinfo {author} {\bibfnamefont {H.~S.}\ \bibnamefont
  {Gutowsky}},\ }\bibfield  {title} {\enquote {\bibinfo {title} {Proton
  magnetic resonance of the $\mbox{CH}_3$ group. {V.} temperature dependence of
  $t_1$ in several molecular crystals},}\ }\href@noop {} {\bibfield  {journal}
  {\bibinfo  {journal} {J. Chem. Phys.}\ }\textbf {\bibinfo {volume} {31}},\
  \bibinfo {pages} {55--65} (\bibinfo {year} {1959})}\BibitemShut {NoStop}%
\bibitem [{\citenamefont {Egelstaff}(1992)}]{Egelstaff}%
  \BibitemOpen
  \bibfield  {author} {\bibinfo {author} {\bibfnamefont {P.~A.}\ \bibnamefont
  {Egelstaff}},\ }\href@noop {} {\emph {\bibinfo {title} {An Introduction to
  the Liquid State}}},\ \bibinfo {edition} {2nd}\ ed.\ (\bibinfo  {publisher}
  {Oxford University Press},\ \bibinfo {address} {Oxford},\ \bibinfo {year}
  {1992})\BibitemShut {NoStop}%
\bibitem [{\citenamefont {Vega}\ and\ \citenamefont
  {Abascal}(2011)}]{vega_2011}%
  \BibitemOpen
  \bibfield  {author} {\bibinfo {author} {\bibfnamefont {C.}~\bibnamefont
  {Vega}}\ and\ \bibinfo {author} {\bibfnamefont {J.~L.~F.}\ \bibnamefont
  {Abascal}},\ }\bibfield  {title} {\enquote {\bibinfo {title} {Simulating
  water with rigid non-polarizable models: a general perspective},}\
  }\href@noop {} {\bibfield  {journal} {\bibinfo  {journal} {Phys. Chem. Chem.
  Phys.}\ }\textbf {\bibinfo {volume} {13}},\ \bibinfo {pages} {19633--19688}
  (\bibinfo {year} {2011})}\BibitemShut {NoStop}%
\bibitem [{\citenamefont {{van der Spoel}}\ \emph {et~al.}(2005)\citenamefont
  {{van der Spoel}}, \citenamefont {Lindahl}, \citenamefont {Hess},
  \citenamefont {Groenhof}, \citenamefont {Mark},\ and\ \citenamefont
  {Berendsen}}]{gromacs4}%
  \BibitemOpen
  \bibfield  {author} {\bibinfo {author} {\bibfnamefont {D.}~\bibnamefont {{van
  der Spoel}}}, \bibinfo {author} {\bibfnamefont {E.}~\bibnamefont {Lindahl}},
  \bibinfo {author} {\bibfnamefont {B.}~\bibnamefont {Hess}}, \bibinfo {author}
  {\bibfnamefont {G.}~\bibnamefont {Groenhof}}, \bibinfo {author}
  {\bibfnamefont {A.~E.}\ \bibnamefont {Mark}}, \ and\ \bibinfo {author}
  {\bibfnamefont {H.~J.~C.}\ \bibnamefont {Berendsen}},\ }\bibfield  {title}
  {\enquote {\bibinfo {title} {{GROMACS}: fast, flexible, and free},}\
  }\href@noop {} {\bibfield  {journal} {\bibinfo  {journal} {J. Comput. Chem.}\
  }\textbf {\bibinfo {volume} {26}},\ \bibinfo {pages} {1701--1718} (\bibinfo
  {year} {2005})}\BibitemShut {NoStop}%
\bibitem [{\citenamefont {Hess}\ \emph {et~al.}(2008)\citenamefont {Hess},
  \citenamefont {Kutzner}, \citenamefont {{van der Spoel}},\ and\ \citenamefont
  {Lindahl}}]{gromacs3}%
  \BibitemOpen
  \bibfield  {author} {\bibinfo {author} {\bibfnamefont {B.}~\bibnamefont
  {Hess}}, \bibinfo {author} {\bibfnamefont {C.}~\bibnamefont {Kutzner}},
  \bibinfo {author} {\bibfnamefont {D.}~\bibnamefont {{van der Spoel}}}, \ and\
  \bibinfo {author} {\bibfnamefont {E.}~\bibnamefont {Lindahl}},\ }\bibfield
  {title} {\enquote {\bibinfo {title} {Gromacs 4: algorithms for highly
  efficient, load-balanced, and scalable molecular simulation},}\ }\href@noop
  {} {\bibfield  {journal} {\bibinfo  {journal} {J. Chem. Theory Comput.}\
  }\textbf {\bibinfo {volume} {4}},\ \bibinfo {pages} {435--447} (\bibinfo
  {year} {2008})}\BibitemShut {NoStop}%
\bibitem [{\citenamefont {S.Nos{\'e}}(1984)}]{Nose:1984}%
  \BibitemOpen
  \bibfield  {author} {\bibinfo {author} {\bibnamefont {S.Nos{\'e}}},\
  }\bibfield  {title} {\enquote {\bibinfo {title} {A molecular dynamics method
  for simulations in the canonical ensemble},}\ }\href@noop {} {\bibfield
  {journal} {\bibinfo  {journal} {Mol. Phys.}\ }\textbf {\bibinfo {volume}
  {52}},\ \bibinfo {pages} {255--268} (\bibinfo {year} {1984})}\BibitemShut
  {NoStop}%
\bibitem [{\citenamefont {Hoover}(1985)}]{Hoover:1985}%
  \BibitemOpen
  \bibfield  {author} {\bibinfo {author} {\bibfnamefont {W.~G.}\ \bibnamefont
  {Hoover}},\ }\bibfield  {title} {\enquote {\bibinfo {title} {Canonical
  dynamics: Equilibrium phase-space distributions},}\ }\href@noop {} {\bibfield
   {journal} {\bibinfo  {journal} {Phys. Rev. A}\ }\textbf {\bibinfo {volume}
  {31}},\ \bibinfo {pages} {1695--1697} (\bibinfo {year} {1985})}\BibitemShut
  {NoStop}%
\bibitem [{\citenamefont {Essmann}\ \emph {et~al.}(1995)\citenamefont
  {Essmann}, \citenamefont {Petera}, \citenamefont {Berkowitz}, \citenamefont
  {Darden}, \citenamefont {Lee},\ and\ \citenamefont
  {Pedersen}}]{Essmann:1995}%
  \BibitemOpen
  \bibfield  {author} {\bibinfo {author} {\bibfnamefont {U.}~\bibnamefont
  {Essmann}}, \bibinfo {author} {\bibfnamefont {L.}~\bibnamefont {Petera}},
  \bibinfo {author} {\bibfnamefont {M.}~\bibnamefont {Berkowitz}}, \bibinfo
  {author} {\bibfnamefont {T.}~\bibnamefont {Darden}}, \bibinfo {author}
  {\bibfnamefont {H.}~\bibnamefont {Lee}}, \ and\ \bibinfo {author}
  {\bibfnamefont {L.}~\bibnamefont {Pedersen}},\ }\bibfield  {title} {\enquote
  {\bibinfo {title} {A smooth particle mesh ewald method},}\ }\href@noop {}
  {\bibfield  {journal} {\bibinfo  {journal} {J. Chem. Phys.}\ }\textbf
  {\bibinfo {volume} {103}},\ \bibinfo {pages} {8577--8593} (\bibinfo {year}
  {1995})}\BibitemShut {NoStop}%
\bibitem [{\citenamefont {Wennberg}\ \emph {et~al.}(2013)\citenamefont
  {Wennberg}, \citenamefont {Murtola}, \citenamefont {Hess},\ and\
  \citenamefont {Lindahl}}]{wennberg_2013}%
  \BibitemOpen
  \bibfield  {author} {\bibinfo {author} {\bibfnamefont {C.~L.}\ \bibnamefont
  {Wennberg}}, \bibinfo {author} {\bibfnamefont {T.}~\bibnamefont {Murtola}},
  \bibinfo {author} {\bibfnamefont {B.}~\bibnamefont {Hess}}, \ and\ \bibinfo
  {author} {\bibfnamefont {E.}~\bibnamefont {Lindahl}},\ }\bibfield  {title}
  {\enquote {\bibinfo {title} {Lennard-jones lattice summation in bilayer
  simulations has critical effects on surface tension and lipid properties},}\
  }\href@noop {} {\bibfield  {journal} {\bibinfo  {journal} {J. Chem. Theory
  Comput.}\ }\textbf {\bibinfo {volume} {9}},\ \bibinfo {pages} {3527--3537}
  (\bibinfo {year} {2013})}\BibitemShut {NoStop}%
\bibitem [{\citenamefont {Wennberg}\ \emph {et~al.}(2015)\citenamefont
  {Wennberg}, \citenamefont {Murtola}, \citenamefont {P\'all}, \citenamefont
  {Abraham}, \citenamefont {Hess},\ and\ \citenamefont
  {Lindahl}}]{wennberg_2015}%
  \BibitemOpen
  \bibfield  {author} {\bibinfo {author} {\bibfnamefont {C.~L.}\ \bibnamefont
  {Wennberg}}, \bibinfo {author} {\bibfnamefont {T.}~\bibnamefont {Murtola}},
  \bibinfo {author} {\bibfnamefont {S.}~\bibnamefont {P\'all}}, \bibinfo
  {author} {\bibfnamefont {M.~J.}\ \bibnamefont {Abraham}}, \bibinfo {author}
  {\bibfnamefont {B.}~\bibnamefont {Hess}}, \ and\ \bibinfo {author}
  {\bibfnamefont {E.}~\bibnamefont {Lindahl}},\ }\bibfield  {title} {\enquote
  {\bibinfo {title} {Direct-space corrections enable fast and accurate
  lorentz-berthelot combination rule lennard-jones lattice summation},}\
  }\href@noop {} {\bibfield  {journal} {\bibinfo  {journal} {J. Chem. Theory
  Comput.}\ }\textbf {\bibinfo {volume} {11}},\ \bibinfo {pages} {5737--5746}
  (\bibinfo {year} {2015})}\BibitemShut {NoStop}%
\bibitem [{\citenamefont {Miyamoto}\ and\ \citenamefont
  {Kollman}(1992)}]{miyamoto_1992}%
  \BibitemOpen
  \bibfield  {author} {\bibinfo {author} {\bibfnamefont {S.}~\bibnamefont
  {Miyamoto}}\ and\ \bibinfo {author} {\bibfnamefont {P.~A.}\ \bibnamefont
  {Kollman}},\ }\bibfield  {title} {\enquote {\bibinfo {title} {Settle: An
  analytical version of the shake and rattle algorithm for rigid water
  models},}\ }\href@noop {} {\bibfield  {journal} {\bibinfo  {journal} {J.
  Comput. Chem.}\ }\textbf {\bibinfo {volume} {13}},\ \bibinfo {pages}
  {952--962} (\bibinfo {year} {1992})}\BibitemShut {NoStop}%
\bibitem [{\citenamefont {Grigera}(2020)}]{grigera_arxiv}%
  \BibitemOpen
  \bibfield  {author} {\bibinfo {author} {\bibfnamefont {T.~S.}\ \bibnamefont
  {Grigera}},\ }\bibfield  {title} {\enquote {\bibinfo {title} {Everything you
  wish to know about correlations but are afraid to ask},}\ }\href@noop {}
  {\bibfield  {journal} {\bibinfo  {journal} {{\tt arXiv:2002.01750v1}}\
  }\textbf {\bibinfo {volume} {{\tt [cond-mat.stat-mech]}}} (\bibinfo {year}
  {2020})}\BibitemShut {NoStop}%
\bibitem [{\citenamefont {Press}\ \emph {et~al.}(1992)\citenamefont {Press},
  \citenamefont {Teukolsky}, \citenamefont {Vetterling},\ and\ \citenamefont
  {Flannery}}]{numrecipes}%
  \BibitemOpen
  \bibfield  {author} {\bibinfo {author} {\bibfnamefont {W.~H.}\ \bibnamefont
  {Press}}, \bibinfo {author} {\bibfnamefont {S.~A.}\ \bibnamefont
  {Teukolsky}}, \bibinfo {author} {\bibfnamefont {W.~T.}\ \bibnamefont
  {Vetterling}}, \ and\ \bibinfo {author} {\bibfnamefont {P.}~\bibnamefont
  {Flannery}},\ }\href@noop {} {\emph {\bibinfo {title} {Numerical Recipes in
  C: The Art of Scientific Computing}}},\ \bibinfo {edition} {2nd}\ ed.\
  (\bibinfo  {publisher} {Cambridge University Press},\ \bibinfo {address}
  {Cambridge, USA},\ \bibinfo {year} {1992})\BibitemShut {NoStop}%
\bibitem [{\citenamefont {Michaud-Agraval}\ \emph {et~al.}(2011)\citenamefont
  {Michaud-Agraval}, \citenamefont {Denning}, \citenamefont {Woolf},\ and\
  \citenamefont {Beckstein}}]{mdanalysis1}%
  \BibitemOpen
  \bibfield  {author} {\bibinfo {author} {\bibfnamefont {N.}~\bibnamefont
  {Michaud-Agraval}}, \bibinfo {author} {\bibfnamefont {E.~J.}\ \bibnamefont
  {Denning}}, \bibinfo {author} {\bibfnamefont {T.~B.}\ \bibnamefont {Woolf}},
  \ and\ \bibinfo {author} {\bibfnamefont {O.}~\bibnamefont {Beckstein}},\
  }\bibfield  {title} {\enquote {\bibinfo {title} {{MDAnalysis}: A toolkit for
  the analysis of molecular dynamics simulations},}\ }\href@noop {} {\bibfield
  {journal} {\bibinfo  {journal} {J. Comput. Chem.}\ }\textbf {\bibinfo
  {volume} {32}},\ \bibinfo {pages} {2319--2327} (\bibinfo {year}
  {2011})}\BibitemShut {NoStop}%
\bibitem [{\citenamefont {Gowers}\ \emph {et~al.}(2016)\citenamefont {Gowers},
  \citenamefont {Linke}, \citenamefont {Barnoud}, \citenamefont {Reddy},
  \citenamefont {Melo}, \citenamefont {Seyler}, \citenamefont {Doma\'nski},
  \citenamefont {Dotson}, \citenamefont {Buchoux}, \citenamefont {Kenney},\
  and\ \citenamefont {Beckstein}}]{mdanalysis2}%
  \BibitemOpen
  \bibfield  {author} {\bibinfo {author} {\bibfnamefont {R.~J.}\ \bibnamefont
  {Gowers}}, \bibinfo {author} {\bibfnamefont {M.}~\bibnamefont {Linke}},
  \bibinfo {author} {\bibfnamefont {J.}~\bibnamefont {Barnoud}}, \bibinfo
  {author} {\bibfnamefont {T.~J.~E.}\ \bibnamefont {Reddy}}, \bibinfo {author}
  {\bibfnamefont {M.~N.}\ \bibnamefont {Melo}}, \bibinfo {author}
  {\bibfnamefont {S.~L.}\ \bibnamefont {Seyler}}, \bibinfo {author}
  {\bibfnamefont {J.}~\bibnamefont {Doma\'nski}}, \bibinfo {author}
  {\bibfnamefont {D.~L.}\ \bibnamefont {Dotson}}, \bibinfo {author}
  {\bibfnamefont {S.}~\bibnamefont {Buchoux}}, \bibinfo {author} {\bibfnamefont
  {I.~M.}\ \bibnamefont {Kenney}}, \ and\ \bibinfo {author} {\bibfnamefont
  {O.}~\bibnamefont {Beckstein}},\ }\bibfield  {title} {\enquote {\bibinfo
  {title} {{MDAnalysis}: A python package for the rapid analysis of molecular
  dynamics simulations},}\ }in\ \href@noop {} {\emph {\bibinfo {booktitle}
  {Proceedings of the 15th Python in Science Conference}}},\ \bibinfo {editor}
  {edited by\ \bibinfo {editor} {\bibfnamefont {S.}~\bibnamefont {Benthall}}\
  and\ \bibinfo {editor} {\bibfnamefont {S.}~\bibnamefont {Rostrup}}}\
  (\bibinfo {address} {Austin, TX},\ \bibinfo {year} {2016})\ pp.\ \bibinfo
  {pages} {98--105}\BibitemShut {NoStop}%
\bibitem [{\citenamefont {Oliphant}(06  )}]{numpy}%
  \BibitemOpen
  \bibfield  {author} {\bibinfo {author} {\bibfnamefont {T.}~\bibnamefont
  {Oliphant}},\ }\href {http://www.numpy.org/} {\enquote {\bibinfo {title}
  {{NumPy}: A guide to {NumPy}},}\ }\bibinfo {howpublished} {USA: Trelgol
  Publishing} (\bibinfo {year} {2006--})\BibitemShut {NoStop}%
\bibitem [{\citenamefont {Virtanen}\ \emph {et~al.}(2020)\citenamefont
  {Virtanen}, \citenamefont {Gommers}, \citenamefont {Oliphant}, \citenamefont
  {Haberland}, \citenamefont {Reddy}, \citenamefont {Cournapeau}, \citenamefont
  {Burovski}, \citenamefont {Peterson}, \citenamefont {Weckesser},
  \citenamefont {Bright}, \citenamefont {{van der Walt}}, \citenamefont
  {Brett}, \citenamefont {Wilson}, \citenamefont {Millman}, \citenamefont
  {Mayorov}, \citenamefont {Nelson}, \citenamefont {Jones}, \citenamefont
  {Kern}, \citenamefont {Larson}, \citenamefont {Carey}, \citenamefont {Polat},
  \citenamefont {Feng}, \citenamefont {Moore}, \citenamefont {{VanderPlas}},
  \citenamefont {Laxalde}, \citenamefont {Perktold}, \citenamefont {Cimrman},
  \citenamefont {Henriksen}, \citenamefont {Quintero}, \citenamefont {Harris},
  \citenamefont {Archibald}, \citenamefont {Ribeiro}, \citenamefont
  {Pedregosa}, \citenamefont {{van Mulbregt}},\ and\ \citenamefont {{SciPy 1.0
  Contributors}}}]{scipy}%
  \BibitemOpen
  \bibfield  {author} {\bibinfo {author} {\bibfnamefont {P.}~\bibnamefont
  {Virtanen}}, \bibinfo {author} {\bibfnamefont {R.}~\bibnamefont {Gommers}},
  \bibinfo {author} {\bibfnamefont {T.~E.}\ \bibnamefont {Oliphant}}, \bibinfo
  {author} {\bibfnamefont {M.}~\bibnamefont {Haberland}}, \bibinfo {author}
  {\bibfnamefont {T.}~\bibnamefont {Reddy}}, \bibinfo {author} {\bibfnamefont
  {D.}~\bibnamefont {Cournapeau}}, \bibinfo {author} {\bibfnamefont
  {E.}~\bibnamefont {Burovski}}, \bibinfo {author} {\bibfnamefont
  {P.}~\bibnamefont {Peterson}}, \bibinfo {author} {\bibfnamefont
  {W.}~\bibnamefont {Weckesser}}, \bibinfo {author} {\bibfnamefont
  {J.}~\bibnamefont {Bright}}, \bibinfo {author} {\bibfnamefont {S.~J.}\
  \bibnamefont {{van der Walt}}}, \bibinfo {author} {\bibfnamefont
  {M.}~\bibnamefont {Brett}}, \bibinfo {author} {\bibfnamefont
  {J.}~\bibnamefont {Wilson}}, \bibinfo {author} {\bibfnamefont {K.~J.}\
  \bibnamefont {Millman}}, \bibinfo {author} {\bibfnamefont {N.}~\bibnamefont
  {Mayorov}}, \bibinfo {author} {\bibfnamefont {A.~R.~J.}\ \bibnamefont
  {Nelson}}, \bibinfo {author} {\bibfnamefont {E.}~\bibnamefont {Jones}},
  \bibinfo {author} {\bibfnamefont {R.}~\bibnamefont {Kern}}, \bibinfo {author}
  {\bibfnamefont {E.}~\bibnamefont {Larson}}, \bibinfo {author} {\bibfnamefont
  {C.~J.}\ \bibnamefont {Carey}}, \bibinfo {author} {\bibfnamefont
  {{\.I}.}~\bibnamefont {Polat}}, \bibinfo {author} {\bibfnamefont
  {Y.}~\bibnamefont {Feng}}, \bibinfo {author} {\bibfnamefont {E.~W.}\
  \bibnamefont {Moore}}, \bibinfo {author} {\bibfnamefont {J.}~\bibnamefont
  {{VanderPlas}}}, \bibinfo {author} {\bibfnamefont {D.}~\bibnamefont
  {Laxalde}}, \bibinfo {author} {\bibfnamefont {J.}~\bibnamefont {Perktold}},
  \bibinfo {author} {\bibfnamefont {R.}~\bibnamefont {Cimrman}}, \bibinfo
  {author} {\bibfnamefont {I.}~\bibnamefont {Henriksen}}, \bibinfo {author}
  {\bibfnamefont {E.~A.}\ \bibnamefont {Quintero}}, \bibinfo {author}
  {\bibfnamefont {C.~R.}\ \bibnamefont {Harris}}, \bibinfo {author}
  {\bibfnamefont {A.~M.}\ \bibnamefont {Archibald}}, \bibinfo {author}
  {\bibfnamefont {A.~H.}\ \bibnamefont {Ribeiro}}, \bibinfo {author}
  {\bibfnamefont {F.}~\bibnamefont {Pedregosa}}, \bibinfo {author}
  {\bibfnamefont {P.}~\bibnamefont {{van Mulbregt}}}, \ and\ \bibinfo {author}
  {\bibnamefont {{SciPy 1.0 Contributors}}},\ }\bibfield  {title} {\enquote
  {\bibinfo {title} {{{SciPy} 1.0: Fundamental Algorithms for Scientific
  Computing in Python}},}\ }\href {\doibase 10.1038/s41592-019-0686-2}
  {\bibfield  {journal} {\bibinfo  {journal} {Nature Methods}\ }\textbf
  {\bibinfo {volume} {17}},\ \bibinfo {pages} {261--272} (\bibinfo {year}
  {2020})}\BibitemShut {NoStop}%
\bibitem [{\citenamefont {Busch}, \citenamefont {Neumann},\ and\ \citenamefont
  {Paschek}(2021)}]{Busch_2021}%
  \BibitemOpen
  \bibfield  {author} {\bibinfo {author} {\bibfnamefont {J.}~\bibnamefont
  {Busch}}, \bibinfo {author} {\bibfnamefont {J.}~\bibnamefont {Neumann}}, \
  and\ \bibinfo {author} {\bibfnamefont {D.}~\bibnamefont {Paschek}},\
  }\bibfield  {title} {\enquote {\bibinfo {title} {An exact a posteriori
  correction for hydrogen bond population correlation functions and other
  reversible geminate recombinations obtained from simulations with periodic
  boundary conditions. liquid water as a test case},}\ }\href@noop {}
  {\bibfield  {journal} {\bibinfo  {journal} {J. Chem. Phys.}\ }\textbf
  {\bibinfo {volume} {154}},\ \bibinfo {pages} {214501} (\bibinfo {year}
  {2021})}\BibitemShut {NoStop}%
\bibitem [{\citenamefont {Allen}\ and\ \citenamefont
  {Tildesley}(1987)}]{allentildesley}%
  \BibitemOpen
  \bibfield  {author} {\bibinfo {author} {\bibfnamefont {M.~P.}\ \bibnamefont
  {Allen}}\ and\ \bibinfo {author} {\bibfnamefont {D.~J.}\ \bibnamefont
  {Tildesley}},\ }\href@noop {} {\emph {\bibinfo {title} {Computer Simulation
  of Liquids}}}\ (\bibinfo  {publisher} {Oxford University Press, Clarendon,
  Oxford},\ \bibinfo {year} {1987})\BibitemShut {NoStop}%
\bibitem [{\citenamefont {Gonz{\'a}les}\ and\ \citenamefont
  {Abascal}(2010)}]{Gonzales_2010}%
  \BibitemOpen
  \bibfield  {author} {\bibinfo {author} {\bibfnamefont {M.}~\bibnamefont
  {Gonz{\'a}les}}\ and\ \bibinfo {author} {\bibfnamefont {J.}~\bibnamefont
  {Abascal}},\ }\bibfield  {title} {\enquote {\bibinfo {title} {The shear
  viscosity of rigid water models},}\ }\href@noop {} {\bibfield  {journal}
  {\bibinfo  {journal} {J. Chem. Phys.}\ }\textbf {\bibinfo {volume} {132}},\
  \bibinfo {pages} {096101} (\bibinfo {year} {2010})}\BibitemShut {NoStop}%
\bibitem [{\citenamefont {Krynicki}(1966)}]{Krynicki_1966}%
  \BibitemOpen
  \bibfield  {author} {\bibinfo {author} {\bibfnamefont {K.}~\bibnamefont
  {Krynicki}},\ }\bibfield  {title} {\enquote {\bibinfo {title} {Proton
  spin-lattice relaxation in pure water between 0$^\circ${C} and
  110$^\circ${C}},}\ }\href@noop {} {\bibfield  {journal} {\bibinfo  {journal}
  {Physica}\ }\textbf {\bibinfo {volume} {32}},\ \bibinfo {pages} {167--178}
  (\bibinfo {year} {1966})}\BibitemShut {NoStop}%
\bibitem [{\citenamefont {v.~Goldammer}\ and\ \citenamefont
  {Zeidler}(1969)}]{Goldammer_1969}%
  \BibitemOpen
  \bibfield  {author} {\bibinfo {author} {\bibfnamefont {E.}~\bibnamefont
  {v.~Goldammer}}\ and\ \bibinfo {author} {\bibfnamefont {M.~D.}\ \bibnamefont
  {Zeidler}},\ }\bibfield  {title} {\enquote {\bibinfo {title} {Molecular
  motion in aqueous mixtures with organic liquids by nmr relaxation
  measurements},}\ }\href@noop {} {\bibfield  {journal} {\bibinfo  {journal}
  {Ber. Bunsenges. Phys. Chem.}\ }\textbf {\bibinfo {volume} {73}},\ \bibinfo
  {pages} {4--15} (\bibinfo {year} {1969})}\BibitemShut {NoStop}%
\bibitem [{\citenamefont {D{\"u}nweg}\ and\ \citenamefont
  {Kremer}(1993)}]{duenweg_1993}%
  \BibitemOpen
  \bibfield  {author} {\bibinfo {author} {\bibfnamefont {B.}~\bibnamefont
  {D{\"u}nweg}}\ and\ \bibinfo {author} {\bibfnamefont {K.}~\bibnamefont
  {Kremer}},\ }\bibfield  {title} {\enquote {\bibinfo {title} {Molecular
  dynamics simulation of a polymer chain in solution},}\ }\href@noop {}
  {\bibfield  {journal} {\bibinfo  {journal} {J. Chem. Phys.}\ }\textbf
  {\bibinfo {volume} {99}},\ \bibinfo {pages} {6983--6997} (\bibinfo {year}
  {1993})}\BibitemShut {NoStop}%
\bibitem [{\citenamefont {Beenakker}(1986)}]{beenacker_1986}%
  \BibitemOpen
  \bibfield  {author} {\bibinfo {author} {\bibfnamefont {C.~W.~J.}\
  \bibnamefont {Beenakker}},\ }\bibfield  {title} {\enquote {\bibinfo {title}
  {Ewald sum of the {Rotne}-{Prager} tensor},}\ }\href@noop {} {\bibfield
  {journal} {\bibinfo  {journal} {J. Chem. Phys.}\ }\textbf {\bibinfo {volume}
  {85}},\ \bibinfo {pages} {1581--1582} (\bibinfo {year} {1986})}\BibitemShut
  {NoStop}%
\bibitem [{\citenamefont {Hasimoto}(1959)}]{hasimoto_1959}%
  \BibitemOpen
  \bibfield  {author} {\bibinfo {author} {\bibfnamefont {H.}~\bibnamefont
  {Hasimoto}},\ }\bibfield  {title} {\enquote {\bibinfo {title} {On the
  periodic fundamental solutions of the stokes equations and their application
  to viscous flow past a cubic array of spheres},}\ }\href@noop {} {\bibfield
  {journal} {\bibinfo  {journal} {J. Fluid Mech.}\ }\textbf {\bibinfo {volume}
  {5}},\ \bibinfo {pages} {317--328} (\bibinfo {year} {1959})}\BibitemShut
  {NoStop}%
\bibitem [{\citenamefont {Laage}\ and\ \citenamefont
  {Hynes}(2006)}]{laage_2006}%
  \BibitemOpen
  \bibfield  {author} {\bibinfo {author} {\bibfnamefont {D.}~\bibnamefont
  {Laage}}\ and\ \bibinfo {author} {\bibfnamefont {J.~T.}\ \bibnamefont
  {Hynes}},\ }\bibfield  {title} {\enquote {\bibinfo {title} {A molecular jump
  mechanism of water reorientation},}\ }\href@noop {} {\bibfield  {journal}
  {\bibinfo  {journal} {Science}\ }\textbf {\bibinfo {volume} {311}},\ \bibinfo
  {pages} {832--835} (\bibinfo {year} {2006})}\BibitemShut {NoStop}%
\bibitem [{\citenamefont {Laage}\ and\ \citenamefont
  {Hynes}(2008)}]{laage_2008}%
  \BibitemOpen
  \bibfield  {author} {\bibinfo {author} {\bibfnamefont {D.}~\bibnamefont
  {Laage}}\ and\ \bibinfo {author} {\bibfnamefont {J.~T.}\ \bibnamefont
  {Hynes}},\ }\bibfield  {title} {\enquote {\bibinfo {title} {On the molecular
  mechanism of water reorientation},}\ }\href@noop {} {\bibfield  {journal}
  {\bibinfo  {journal} {J. Phys. Chem. B}\ }\textbf {\bibinfo {volume} {112}},\
  \bibinfo {pages} {14230--14242} (\bibinfo {year} {2008})}\BibitemShut
  {NoStop}%
\bibitem [{\citenamefont {Celebi}\ \emph {et~al.}(2021)\citenamefont {Celebi},
  \citenamefont {Jamali}, \citenamefont {Bardow}, \citenamefont {Vlugt},\ and\
  \citenamefont {Moultos}}]{celebi_2021}%
  \BibitemOpen
  \bibfield  {author} {\bibinfo {author} {\bibfnamefont {A.~T.}\ \bibnamefont
  {Celebi}}, \bibinfo {author} {\bibfnamefont {S.~H.}\ \bibnamefont {Jamali}},
  \bibinfo {author} {\bibfnamefont {A.}~\bibnamefont {Bardow}}, \bibinfo
  {author} {\bibfnamefont {T.~J.~H.}\ \bibnamefont {Vlugt}}, \ and\ \bibinfo
  {author} {\bibfnamefont {O.~A.}\ \bibnamefont {Moultos}},\ }\bibfield
  {title} {\enquote {\bibinfo {title} {Finite-size effects of diffusion
  coefficients computed from molecular dynamics: a review of what we have
  learned so far},}\ }\href@noop {} {\bibfield  {journal} {\bibinfo  {journal}
  {Mol. Simul.}\ }\textbf {\bibinfo {volume} {47}},\ \bibinfo {pages}
  {831--845} (\bibinfo {year} {2021})}\BibitemShut {NoStop}%
\bibitem [{\citenamefont {Krynicki}, \citenamefont {Green},\ and\ \citenamefont
  {Sawyer}(1978)}]{krynicki_1978}%
  \BibitemOpen
  \bibfield  {author} {\bibinfo {author} {\bibfnamefont {K.}~\bibnamefont
  {Krynicki}}, \bibinfo {author} {\bibfnamefont {C.~D.}\ \bibnamefont {Green}},
  \ and\ \bibinfo {author} {\bibfnamefont {D.~W.}\ \bibnamefont {Sawyer}},\
  }\bibfield  {title} {\enquote {\bibinfo {title} {Pressure and temperature
  dependence of self-diffusion in water},}\ }\href@noop {} {\bibfield
  {journal} {\bibinfo  {journal} {Faraday Discuss. Chem. Soc.}\ }\textbf
  {\bibinfo {volume} {66}},\ \bibinfo {pages} {199--208} (\bibinfo {year}
  {1978})}\BibitemShut {NoStop}%
\end{thebibliography}%

\end{document}